\documentclass{aastex631}
\usepackage{rotating}
\usepackage{amsmath}

\newcommand{\Autoref}[1]{%
  \begingroup%
  \def\sectionautorefname{Section}%
  \def\tableautorefname{Table}%
  \def\equationautorefname{Equation}%
  \autoref{#1}%
  \endgroup%
}

\let\tablenum\relax
\usepackage[range-phrase=\textendash, range-units=single, separate-uncertainty=true]{siunitx}
\let\sun\odot
\DeclareSIUnit\solarmass{\ensuremath{M_\sun}}
\DeclareSIUnit\erg{erg}
\DeclareSIUnit\mwe{m.w.e.}
\DeclareSIUnit\year{yr}
\DeclareSIUnit\parsec{pc}

\renewcommand{\vec}[1]{\boldsymbol{#1}}

\received{April 20, 2021}
\revised{June 15, 2021}
\accepted{June 16, 2021}
\published{September 13, 2021}
\submitjournal{ApJ}

\shorttitle{Super-Kamiokande GW Follow-up}
\shortauthors{The Super-Kamiokande collaboration}

\graphicspath{{./}{figures/}}

\begin{document}

\title{Search for neutrinos in coincidence with gravitational wave events \\ from the LIGO-Virgo O3a Observing Run with the Super-Kamiokande detector}

\newcommand{\AFFicrr}{\affiliation{Kamioka Observatory, Institute for Cosmic Ray Research, University of Tokyo, Kamioka, Gifu 506-1205, Japan}}
\newcommand{\AFFkashiwa}{\affiliation{Research Center for Cosmic Neutrinos, Institute for Cosmic Ray Research, University of Tokyo, Kashiwa, Chiba 277-8582, Japan}}
\newcommand{\AFFicrronly}{\affiliation{Institute for Cosmic Ray Research, University of Tokyo, Kashiwa, Chiba 277-8582, Japan}}
\newcommand{\AFFipmu}{\affiliation{Kavli Institute for the Physics and Mathematics of the Universe (WPI), The University of Tokyo Institutes for Advanced Study, University of Tokyo, Kashiwa, Chiba 277-8583, Japan }}
\newcommand{\AFFmad}{\affiliation{Department of Theoretical Physics, University Autonoma Madrid, 28049 Madrid, Spain}}
\newcommand{\AFFubc}{\affiliation{Department of Physics and Astronomy, University of British Columbia, Vancouver, BC, V6T1Z4, Canada}}
\newcommand{\AFFbu}{\affiliation{Department of Physics, Boston University, Boston, MA 02215, USA}}
\newcommand{\AFFuci}{\affiliation{Department of Physics and Astronomy, University of California, Irvine, Irvine, CA 92697-4575, USA }}
\newcommand{\AFFcsu}{\affiliation{Department of Physics, California State University, Dominguez Hills, Carson, CA 90747, USA}}
\newcommand{\AFFcnm}{\affiliation{Institute for Universe and Elementary Particles, Chonnam National University, Gwangju 61186, Korea}}
\newcommand{\AFFduke}{\affiliation{Department of Physics, Duke University, Durham NC 27708, USA}}
\newcommand{\AFFfukuoka}{\affiliation{Junior College, Fukuoka Institute of Technology, Fukuoka, Fukuoka 811-0295, Japan}}
\newcommand{\AFFgifu}{\affiliation{Department of Physics, Gifu University, Gifu, Gifu 501-1193, Japan}}
\newcommand{\AFFgist}{\affiliation{GIST College, Gwangju Institute of Science and Technology, Gwangju 500-712, Korea}}
\newcommand{\AFFuh}{\affiliation{Department of Physics and Astronomy, University of Hawaii, Honolulu, HI 96822, USA}}
\newcommand{\AFFicl}{\affiliation{Department of Physics, Imperial College London , London, SW7 2AZ, United Kingdom }}
\newcommand{\AFFkek}{\affiliation{High Energy Accelerator Research Organization (KEK), Tsukuba, Ibaraki 305-0801, Japan }}
\newcommand{\AFFkobe}{\affiliation{Department of Physics, Kobe University, Kobe, Hyogo 657-8501, Japan}}
\newcommand{\AFFkyoto}{\affiliation{Department of Physics, Kyoto University, Kyoto, Kyoto 606-8502, Japan}}
\newcommand{\AFFliv}{\affiliation{Department of Physics, University of Liverpool, Liverpool, L69 7ZE, United Kingdom}}
\newcommand{\AFFmiyagi}{\affiliation{Department of Physics, Miyagi University of Education, Sendai, Miyagi 980-0845, Japan}}
\newcommand{\AFFnagoya}{\affiliation{Institute for Space-Earth Environmental Research, Nagoya University, Nagoya, Aichi 464-8602, Japan}}
\newcommand{\AFFkmi}{\affiliation{Kobayashi-Maskawa Institute for the Origin of Particles and the Universe, Nagoya University, Nagoya, Aichi 464-8602, Japan}}
\newcommand{\AFFpol}{\affiliation{National Centre For Nuclear Research, 02-093 Warsaw, Poland}}
\newcommand{\AFFsuny}{\affiliation{Department of Physics and Astronomy, State University of New York at Stony Brook, NY 11794-3800, USA}}
\newcommand{\AFFokayama}{\affiliation{Department of Physics, Okayama University, Okayama, Okayama 700-8530, Japan }}
\newcommand{\AFFosaka}{\affiliation{Department of Physics, Osaka University, Toyonaka, Osaka 560-0043, Japan}}
\newcommand{\AFFox}{\affiliation{Department of Physics, Oxford University, Oxford, OX1 3PU, United Kingdom}}
\newcommand{\AFFqmul}{\affiliation{School of Physics and Astronomy, Queen Mary University of London, London, E1 4NS, United Kingdom}}
\newcommand{\AFFregina}{\affiliation{Department of Physics, University of Regina, 3737 Wascana Parkway, Regina, SK, S4SOA2, Canada}}
\newcommand{\AFFseoul}{\affiliation{Department of Physics, Seoul National University, Seoul 151-742, Korea}}
\newcommand{\AFFsheff}{\affiliation{Department of Physics and Astronomy, University of Sheffield, S3 7RH, Sheffield, United Kingdom}}
\newcommand{\AFFshizuokasc}{\affiliation{Department of Informatics in Social Welfare, Shizuoka University of Welfare, Yaizu, Shizuoka, 425-8611, Japan}}
\newcommand{\AFFstfc}{\affiliation{STFC, Rutherford Appleton Laboratory, Harwell Oxford, and Daresbury Laboratory, Warrington, OX11 0QX, United Kingdom}}
\newcommand{\AFFskk}{\affiliation{Department of Physics, Sungkyunkwan University, Suwon 440-746, Korea}}
\newcommand{\AFFtokyo}{\affiliation{The University of Tokyo, Bunkyo, Tokyo 113-0033, Japan }}
\newcommand{\AFFtodai}{\affiliation{Department of Physics, University of Tokyo, Bunkyo, Tokyo 113-0033, Japan }}
\newcommand{\AFFtit}{\affiliation{Department of Physics,Tokyo Institute of Technology, Meguro, Tokyo 152-8551, Japan }}
\newcommand{\AFFtus}{\affiliation{Department of Physics, Faculty of Science and Technology, Tokyo University of Science, Noda, Chiba 278-8510, Japan }}
\newcommand{\AFFtoronto}{\affiliation{Department of Physics, University of Toronto, ON, M5S 1A7, Canada }}
\newcommand{\AFFtriumf}{\affiliation{TRIUMF, 4004 Wesbrook Mall, Vancouver, BC, V6T2A3, Canada }}
\newcommand{\AFFtokai}{\affiliation{Department of Physics, Tokai University, Hiratsuka, Kanagawa 259-1292, Japan}}
\newcommand{\AFFtsinghua}{\affiliation{Department of Engineering Physics, Tsinghua University, Beijing, 100084, China}}
\newcommand{\AFFynu}{\affiliation{Department of Physics, Yokohama National University, Yokohama, Kanagawa, 240-8501, Japan}}
\newcommand{\AFFllr}{\affiliation{Ecole Polytechnique, IN2P3-CNRS, Laboratoire Leprince-Ringuet, F-91120 Palaiseau, France }}
\newcommand{\AFFbari}{\affiliation{Dipartimento Interuniversitario di Fisica, INFN Sezione di Bari and Universit\`a e Politecnico di Bari, I-70125, Bari, Italy}}
\newcommand{\AFFnapoli}{\affiliation{Dipartimento di Fisica, INFN Sezione di Napoli and Universit\`a di Napoli, I-80126, Napoli, Italy}}
\newcommand{\AFFroma}{\affiliation{INFN Sezione di Roma and Universit\`a di Roma ``La Sapienza'', I-00185, Roma, Italy}}
\newcommand{\AFFpadova}{\affiliation{Dipartimento di Fisica, INFN Sezione di Padova and Universit\`a di Padova, I-35131, Padova, Italy}}
\newcommand{\AFFkeio}{\affiliation{Department of Physics, Keio University, Yokohama, Kanagawa, 223-8522, Japan}}
\newcommand{\AFFwinnipeg}{\affiliation{Department of Physics, University of Winnipeg, MB R3J 3L8, Canada}}
\newcommand{\AFFkcl}{\affiliation{Department of Physics, King's College London, London, WC2R 2LS, UK}}
\newcommand{\AFFwarwick}{\affiliation{Department of Physics, University of Warwick, Coventry, CV4 7AL, UK}}
\newcommand{\AFFral}{\affiliation{Rutherford Appleton Laboratory, Harwell, Oxford, OX11 0QX, UK}}
\newcommand{\AFFwu}{\affiliation{Faculty of Physics, University of Warsaw, Warsaw, 02-093, Poland}}
\newcommand{\AFFbcit}{\affiliation{Department of Physics, British Columbia Institute of Technology, Burnaby, BC, V5G 3H2, Canada}}
\newcommand{\AFFtohoku}{\affiliation{Department of Physics, Faculty of Science, Tohoku University, Sendai, Miyagi, 980-8578, Japan}}

\AFFicrr
\AFFkashiwa
\AFFicrronly
\AFFmad
\AFFbu
\AFFbcit
\AFFuci
\AFFcsu
\AFFcnm
\AFFduke
\AFFllr
\AFFfukuoka
\AFFgifu
\AFFgist
\AFFuh
\AFFicl
\AFFbari
\AFFnapoli
\AFFpadova
\AFFroma
\AFFkeio
\AFFkek
\AFFkcl
\AFFkobe
\AFFkyoto
\AFFliv
\AFFmiyagi
\AFFnagoya
\AFFkmi
\AFFpol
\AFFsuny
\AFFokayama
\AFFosaka
\AFFox
\AFFral
\AFFseoul
\AFFsheff
\AFFshizuokasc
\AFFstfc
\AFFskk
\AFFtohoku
\AFFtokai
\AFFtokyo
\AFFtodai
\AFFipmu
\AFFtit
\AFFtus
\AFFtoronto
\AFFtriumf
\AFFtsinghua
\AFFwu
\AFFwarwick
\AFFwinnipeg
\AFFynu

\author{K.~Abe}
\AFFicrr
\AFFipmu
\author{C.~Bronner}
\AFFicrr
\author{Y.~Hayato}
\AFFicrr
\AFFipmu
\author{M.~Ikeda}
\author{S.~Imaizumi}
\AFFicrr 
\author{J.~Kameda}
\AFFicrr
\AFFipmu
\author{Y.~Kanemura}
\author{Y.~Kataoka}
\author{S.~Miki}
\AFFicrr
\author{M.~Miura} 
\author[0000-0001-7630-2839]{S.~Moriyama} 
\AFFicrr
\AFFipmu
% ...


\collaboration{1000}{The Super-Kamiokande Collaboration}

\begin{abstract}
The Super-Kamiokande detector can be used to search for neutrinos in time coincidence with gravitational waves detected by the LIGO-Virgo Collaboration (LVC). Both low-energy (\SIrange{7}{100}{\mega\electronvolt}) and high-energy (\SIrange{0.1}{e5}{\giga\electronvolt}) samples were analyzed in order to cover a very wide neutrino spectrum. 
Follow-ups of 36 (out of 39) gravitational waves reported in the GWTC-2 catalog were examined; no significant excess above the background was observed, with 10 (24) observed neutrinos compared with 4.8 (25.0) expected events in the high-energy (low-energy) samples. A statistical approach was used to compute the significance of potential coincidences. For each observation, p-values were estimated using neutrino direction and LVC sky map ; the most significant event (GW190602\_175927) is associated with a post-trial p-value of $7.8\%$ ($1.4\sigma$). Additionally, flux limits were computed independently for each sample, and by combining the samples. The energy emitted as neutrinos by the identified gravitational wave sources was constrained, both for given flavors and for all-flavors assuming equipartition between the different flavors, independently for each trigger and by combining sources of the same nature.
\end{abstract}

\vspace*{1cm}

\section{Introduction}
\label{sec:intro}

We have entered a new phase of astronomical observations, the so-called multimessenger astronomy era. Experiments and observatories are more than ever now able to observe the sky in different energy regions (from eV to EeV) with different messengers (photons, cosmic rays, neutrinos, or gravitational waves).

Since 2019 April, the LIGO/Virgo collaboration (LVC) has been publicly releasing their alerts for gravitational waves (GW) directly through their own GraceDB service and through the GCN system~\citep{GCN}. Within a few minutes after the first detection, the first alert is sent with a precise time stamp and a rough sky localization allowing quick follow-ups from other observatories. More precise information on localization is provided in the following days.

The detected GW emitters are categorized by LVC into several types, for which high-energy neutrino (HE-$\nu$) emission is also expected from relativistic outflows and hadronic interactions within these sources: binary neutron star mergers (BNS,~\cite{Kimura:2018vvz}), neutron star-black hole mergers (NSBH,~\cite{Kimura:2017kan}) or binary black hole mergers (BBH,~\cite{Kotera:2016dmp}). Such astrophysical objects may also emit \si{\mega\electronvolt} neutrinos (LE-$\nu$) (for BNS, see \cite{Foucart:2015gaa}). 

However, such a joint observation of GWs and neutrinos is yet to be observed. Even a single event of this type would provide useful information on the underlying mechanisms. Furthermore, high-energy neutrinos would allow improving the localization in the sky of a single GW event, increasing the chance for a pointing observatory (e.g. follow-up telescopes) to observe a third correlated signal if the alert is provided promptly.

The IceCube~\citep{Countryman:2019pqq} and ANTARES/KM3NeT~\citep{Dornic:2020wwv} experiments are already taking part in such follow-up program for every single reported GW event. Nevertheless, these neutrino telescopes are mainly covering HE-$\nu$ above \SI{100}{\giga\electronvolt}. Super-Kamiokande (SK) can complement such follow-up studies, as, despite its much smaller size, it is sensitive to lower energies (\si{\mega\electronvolt}-\si{\tera\electronvolt}). In the past, SK has performed such studies, but only for a few of the detected GW events: GW150914/GW151226 in \cite{Abe:2016jwn} and GW170817 in \cite{Abe:2018mic}. For the \si{\mega\electronvolt} region, searches have also been carried out in KamLAND~\citep{Abe:2020zpn} and Borexino~\citep{Agostini:2017pfa}. 

This paper is focused on the follow-ups of gravitational wave triggers detected during the first half of the third observing run (O3a) of LVC, from 2019 April to 2019 September and described in the GWTC-2 catalog \citep{Abbott:2020niy}. Each GW was classified as a BNS, BBH or NSBH based on the measured masses of the two objects ($m<\SI{3}{\solarmass}$ = Neutron Star, $m>\SI{3}{\solarmass}$ = Black Hole, where \si{\solarmass} is the solar mass). 

This article is organized as follows. \Autoref{sec:sk} describes SK and the used datasets. In \autoref{sec:search}, the search method and basic results will be described. \Autoref{sec:stat} focuses on the extraction of flux limits and signal significance out of each of individual follow-up, while \autoref{sec:pop} describes how the results can be combined to constrain different source populations. \Autoref{sec:concl} summarizes and concludes the discussion. The data release accompanying this article \citep{ZenodoLink} includes all the figures, the tables of observations and calculated flux limits, and the SK effective area.

\section{Super-Kamiokande and event samples}
\label{sec:sk}

SK \citep{Fukuda:2002uc} is a water Cerenkov detector located in the Mozumi mine in Gifu Prefecture, Japan. It lies under Mt. Ikeno (Ikenoyama) with a total of \SI{2700}{\mwe} (meters water equivalent) mean overburden, reducing the cosmic-ray muon rate at the detector by a factor of $\sim 10^5$ with respect to the surface. The detector consists of a cylindrical stainless-steel tank of \SI{39}{\meter} diameter and \SI{42}{\meter} height, filled with \SI{50}{\kilo\tonne} of water. It is optically separated into an inner detector (ID) and an outer detector (OD) by a structure at $\sim \SI{2}{\meter}$ from the wall. The ID is instrumented with 11,129 photomultiplier tubes (PMTs) to observe the Cerenkov light emitted by charged particles produced in neutrino interactions. The OD, instrumented with 1885 PMTs, is primarily used as a veto for external background. SK is sensitive to neutrinos with energies ranging from several \si{\mega\electronvolt} to \si{\tera\electronvolt}.

The experiment has been operating since 1996, and data taking can be separated into six distinct periods, from SK-I to SK-VI, with the latter starting in 2020 July, being the first phase where gadolinium sulfate has been dissolved into the otherwise pure water. In this paper focused on O3a GW events, only data from SK-V (2019 January - 2020 July) were used for analysis, as this is covering the full O3 period.

\subsection{HE-\texorpdfstring{$\nu$}{nu} samples}
\label{sec:sk_HE}

The high-energy samples correspond to neutrinos with $E_\nu > \SI{100}{\mega\electronvolt}$ (which is linked to an electron equivalent energy / visible energy greater than \SI{30}{\mega\electronvolt}). The neutrino is detected thanks to the outgoing lepton produced in the neutrino charged-current interaction. Data are further divided into three sub-samples based on event topology. 

The fully contained (FC) and partially contained (PC) samples have a reconstructed neutrino interaction vertex inside the fiducial volume of the ID\footnote{The fiducial volume for this analysis is defined as the region in the ID more than \SI{1}{\meter} (\SI{2}{\meter} for PC) from any wall.}. The separation between FC and PC is based on the number of effective PMT hits in OD ($<16$ hits for FC, $\geq 16$ hits for PC).

The muons entering the detector can originate from muon neutrino interactions in the rock surrounding the detector. As such events would be indistinguishable from downward-going cosmic-ray muons, only events with upward-going direction are considered, hence the name UPMU (for ``Upward-going muons''). Events are either through-going (with a requirement on track length $>\SI{7}{\meter}$) or stopping in the detector (with a requirement on reconstructed muon momentum $>\SI{1.6}{\giga\electronvolt}$). Further details are documented in~\cite{Ashie:2005ik}.

Typical neutrino energy for FC (PC) will be between \SIlist{0.1;10}{\giga\electronvolt} (\SIlist{1;100}{\giga\electronvolt}) and these samples are sensitive to $\nu_\mu$, $\nu_e$, $\bar\nu_\mu$ and $\bar\nu_e$. The UPMU sample is only sensitive to muon neutrinos and muon antineutrinos, but it covers energies from $\mathcal{O}(\si{\giga\electronvolt})$ to $\mathcal{O}(\si{\tera\electronvolt})$. The contribution of tau neutrinos is subdominant and was therefore neglected in the following, even though it may improve the final limits in a next iteration of the analysis.

\subsection{LE-\texorpdfstring{$\nu$}{nu} sample}
\label{sec:sk_LE}

The low-energy sample corresponds to events with energy between $7$ and $\SI{100}{\mega\electronvolt}$. The largest cross section in this range is the inverse beta decay (IBD) of electron antineutrinos ($\bar\nu_e + p \rightarrow e^{+} + n$) and the second most dominant is neutrino elastic scattering ($\nu + e^{-} \rightarrow \nu + e^{-}$), which is sensitive not only to electron neutrinos but also to other flavors. Interactions on $^{16}\mathrm{O}$ are neglected in the following analysis.

There are two existing data samples in the SK low-energy analysis. In the \SIrange{7}{15.5}{\mega\electronvolt} range, the selection tuned for the solar neutrino analysis~\citep{Abe:2016nxk} is applied, while the supernova relic neutrino (SRN) search~\citep{Bays:2011si} is focused on the \SIrange{15.5}{100}{\mega\electronvolt} range. The main background below \SI{20}{\mega\electronvolt} is spallation products from cosmic-ray muons~\citep{Super-Kamiokande:2015xra} ; above \SI{20}{\mega\electronvolt}, it is dominated by interactions from atmospheric neutrinos and electrons from muon decays.

The solar neutrino analysis is in principle sensitive down to \SI{3.5}{\mega\electronvolt}~\citep{Abe:2016nxk}. However, to reduce the background originating from radioactive decays (especially $\mathrm{Rn}$~\cite{Nakano:2019bnr}), only events with reconstructed energy above \SI{7}{\mega\electronvolt} are considered in this paper.

The expected background is higher than for the HE-$\nu$ samples (see the next section), and, as opposed to the latter, the reconstructed neutrino direction cannot be reliably used to identify spatial coincidence with the GW localization, IBD being mostly insensitive to the original direction.

\section{Search method and results}
\label{sec:search}
 
The information related to O3a GW triggers are extracted from the FITS files \citep{FITS} in the data release accompanying \citep{Abbott:2020niy}. The main input for the SK analysis is the trigger time $t_{\rm GW}$: it is used to define a $\pm \SI{500}{\second}$ time window centered on $t_{\rm GW}$. The choice of this window is based on the conservative considerations proposed by \cite{Baret:2011tk}. The SK data in this window were collected and divided into the four samples (three HE-$\nu$, one LE-$\nu$) described in \autoref{sec:sk}.

Downtime periods, due to detector calibration or other maintenance (e.g., preparation for Gd-loading in early 2020), prevent the follow-up of some GW triggers. Out of the 39 confirmed events from O3a, SK was able to perform the analysis for 36 of them, with a live time within the $\SI{1000}{\second}$ window of $\sim 99.5\%$ for each (not $100\%$ because of cosmic muon vetos and other trigger dead times). Additionally, one of those (GW190512\_180714) was not suited for low-energy analysis since there were large noise fluctuations in SK near the GW time due to high-voltage issues.

\subsection{HE-\texorpdfstring{$\nu$}{nu} samples} 

The events passing the selection described in \autoref{sec:sk_HE} were stored. For FC and PC events, the total visible energy $E_{\rm vis}$ is a good estimator of the incoming neutrino energy. For UPMU events, the reconstructed muon momentum $p_\mu$ is not an accurate estimator because the original neutrino energy cannot be inferred as it interacted in the surrounding rock; it is, however, a lower bound for the neutrino energy\footnote{In the following, we will use the notation $E_{\rm reco}$ to refer to $E_{\rm vis}$ and $p_{\mu}$ for FC/PC and UPMU respectively, with the related caveats.}.

For FC and PC, the direction of each event was estimated by reconstructing the Cerenkov rings in the ID, while the direction of UPMU event is determined using the OD hit information. This local direction was converted to equatorial coordinates, right ascension (\textrm{RA}) and declination (\textrm{Dec}), so that it can easily be compared with GW localization. The associated angular uncertainty was obtained by comparing the reconstructed direction with the true neutrino direction in atmospheric Monte Carlo samples of similar energies. For the lower energies ($E_\nu \lesssim \mathcal{O}(\si{\giga\electronvolt})$), the angular resolution is limited by the scattering angle between the neutrino and the lepton (e.g. $\sigma \sim \SI{20}{\degree}$ for FC, $E_\nu = \SI{2}{\giga\electronvolt}$). Resolution of the order of the degree can be achieved with the UPMU sample, as detailed in~\cite{Hagiwara:2020}.

The expected background rate in the high-energy samples was stable over the full data period and therefore can be extracted from data using the full dataset from 2019 February to 2020 March. The expected number of background events in a 1000-second time window is 0.112, 0.007 and 0.016 events for FC, PC, and UPMU, respectively (with negligible statistical uncertainties).

\begin{figure*}[hbtp]
\gridline{\fig{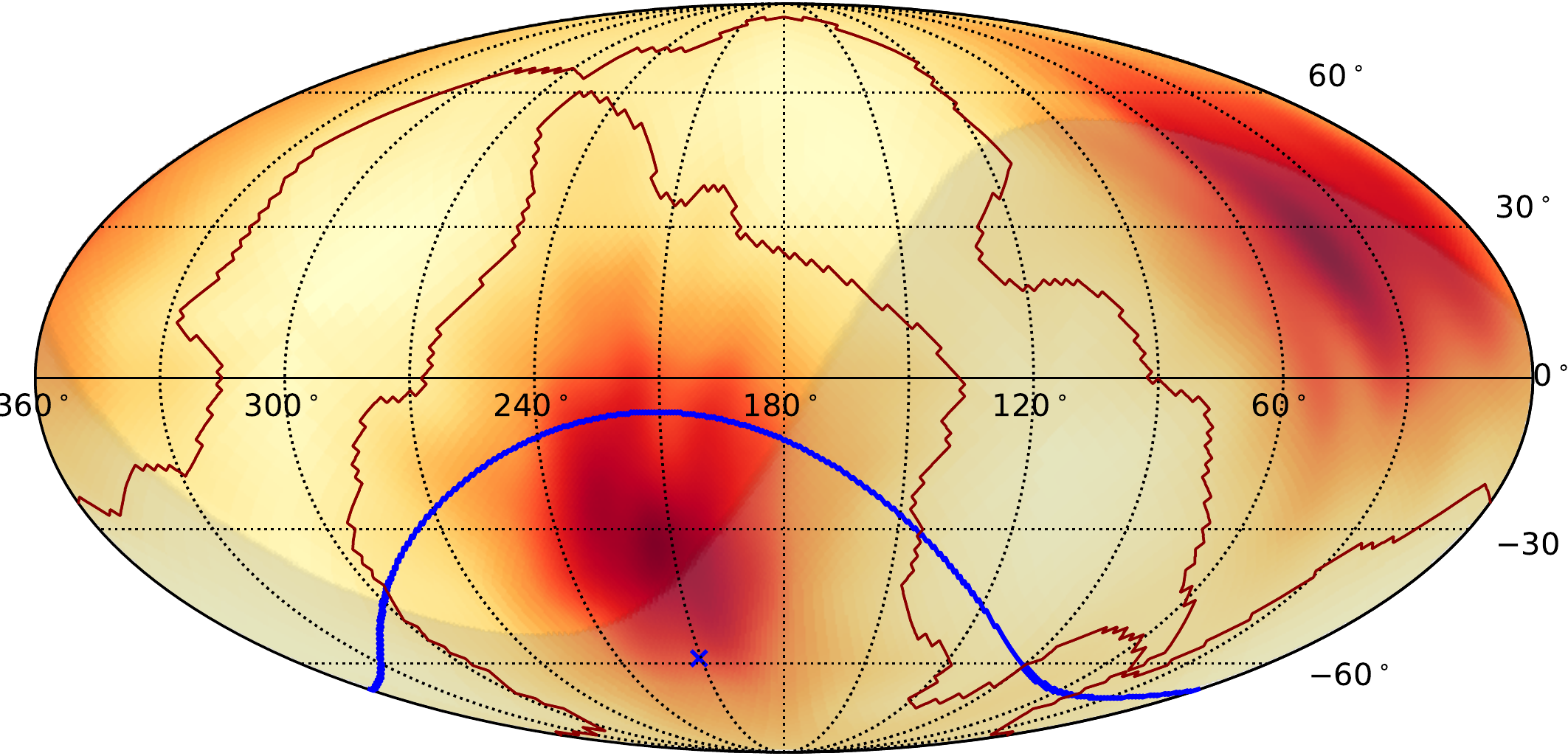}{0.36\textwidth}{(a) GW190424\_180648 (FC)}
          \fig{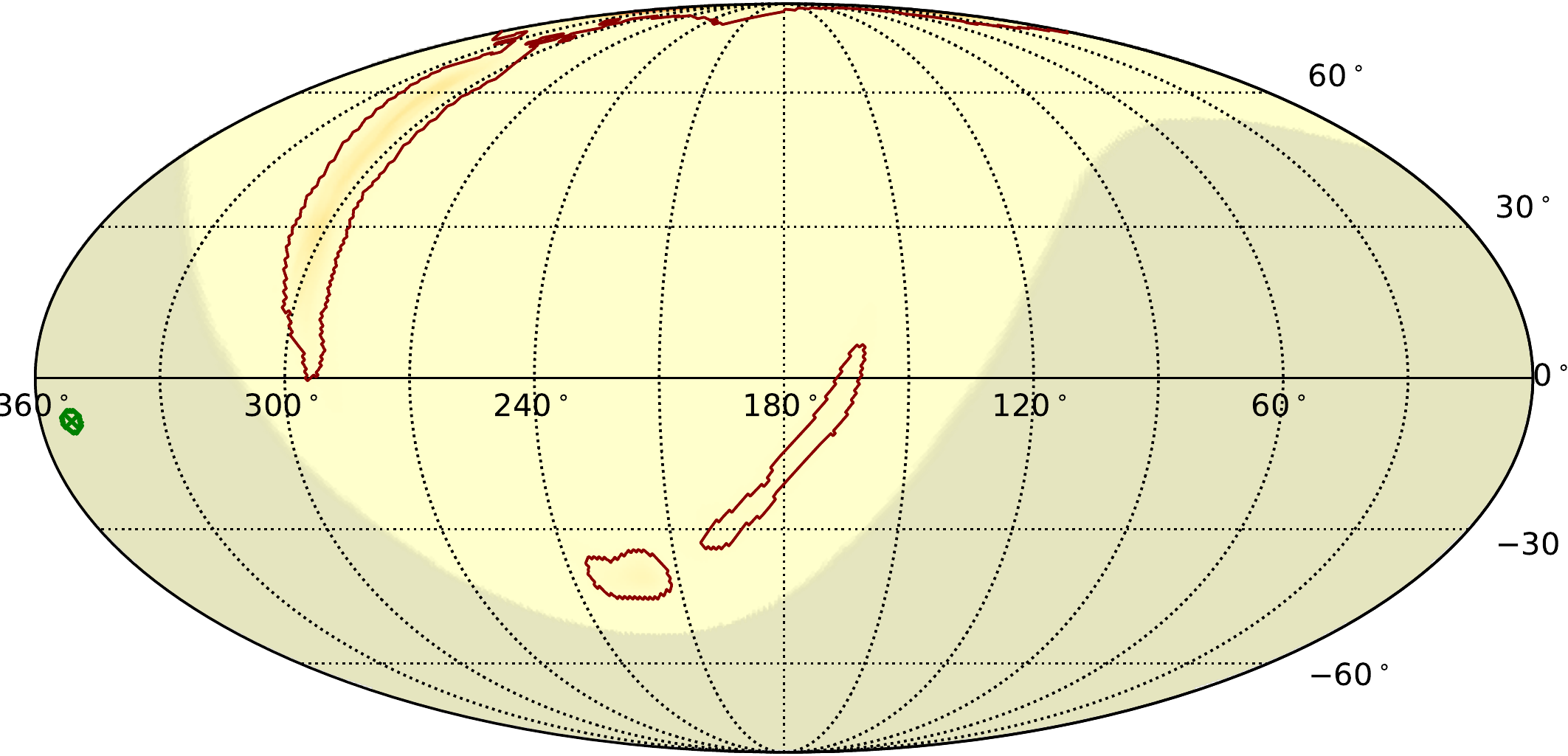}{0.36\textwidth}{(b) GW190426\_152155 (UPMU)}
}
\gridline{
          \fig{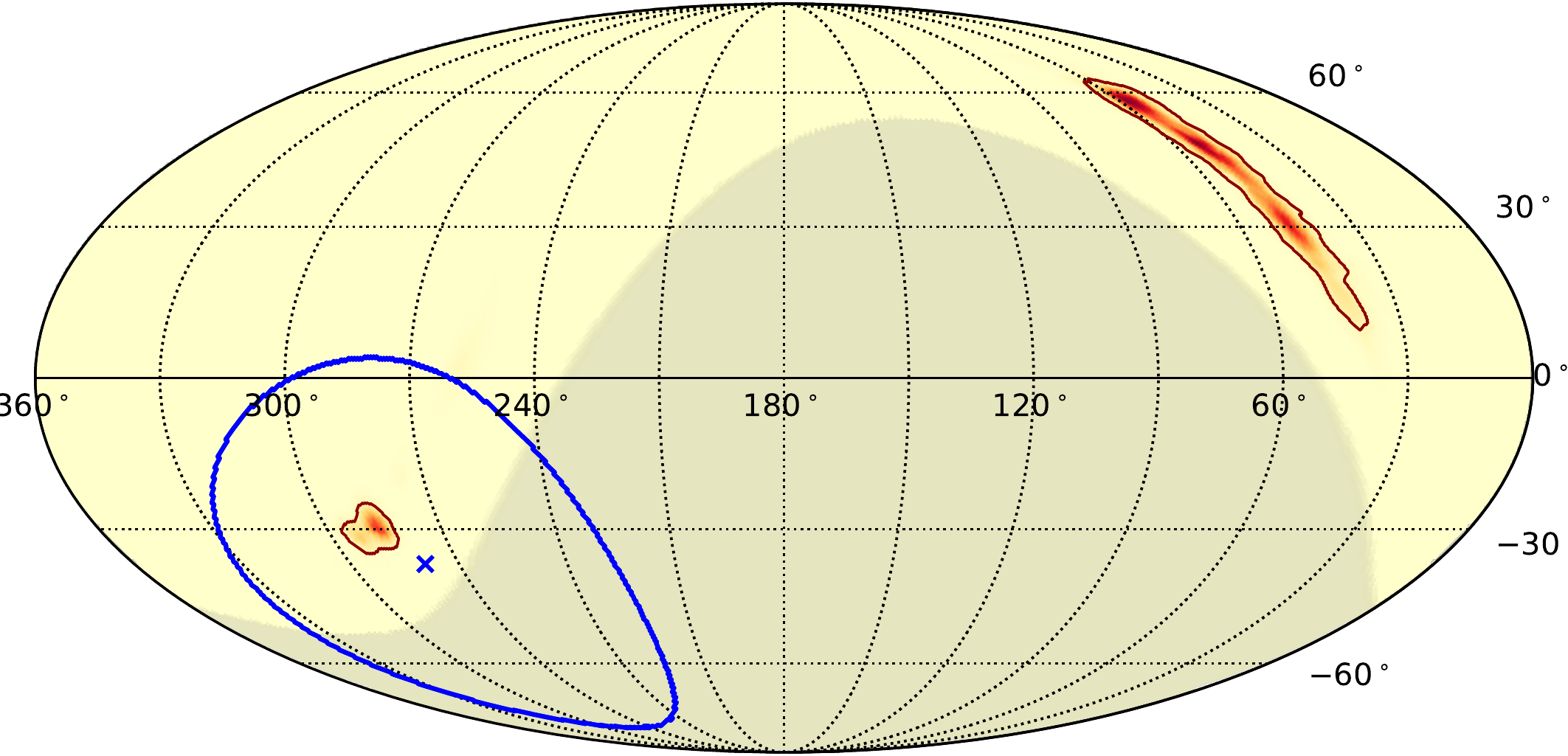}{0.36\textwidth}{(c) GW190513\_205428 (FC)}
          \fig{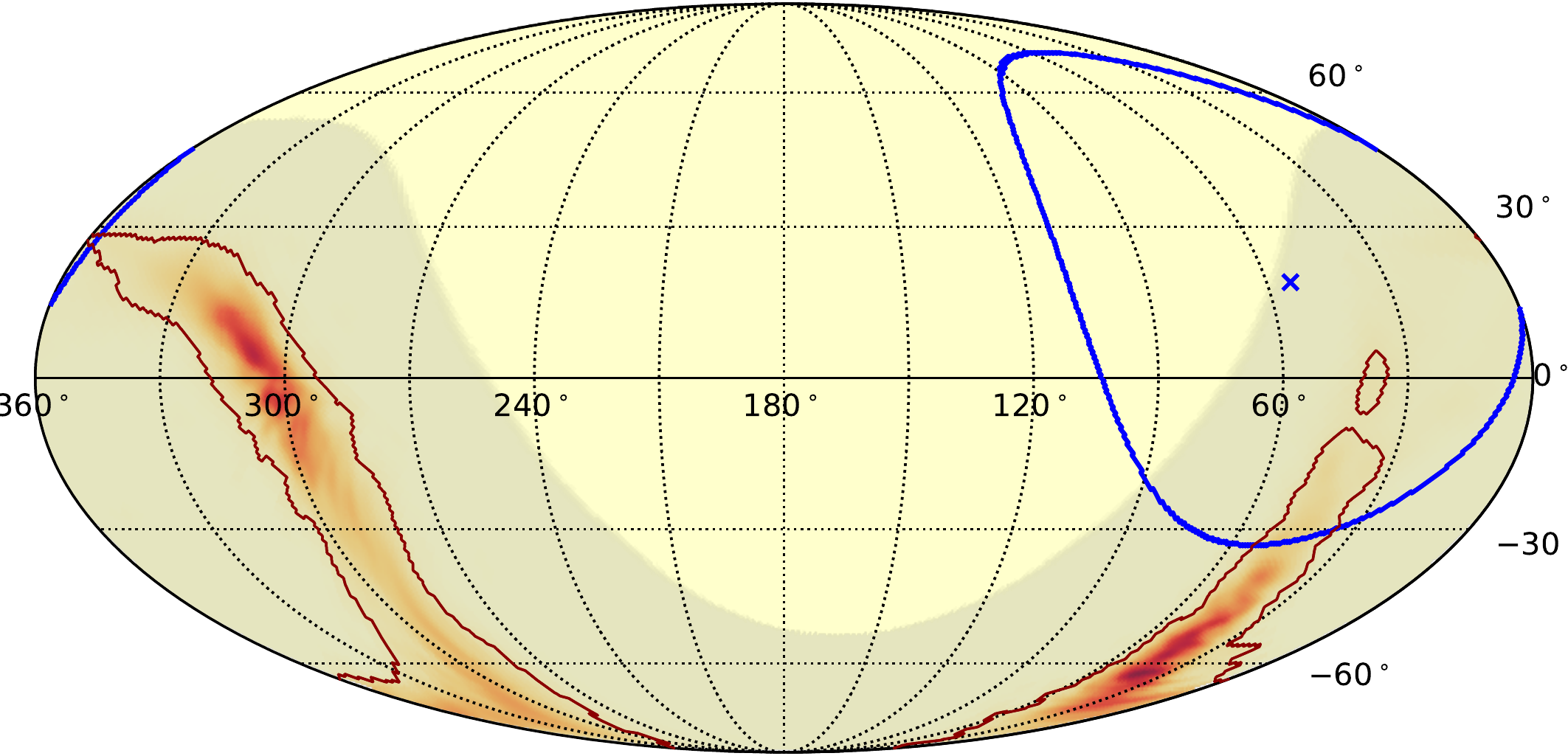}{0.36\textwidth}{(d) GW190527\_092055 (FC)}
}
\gridline{
          \fig{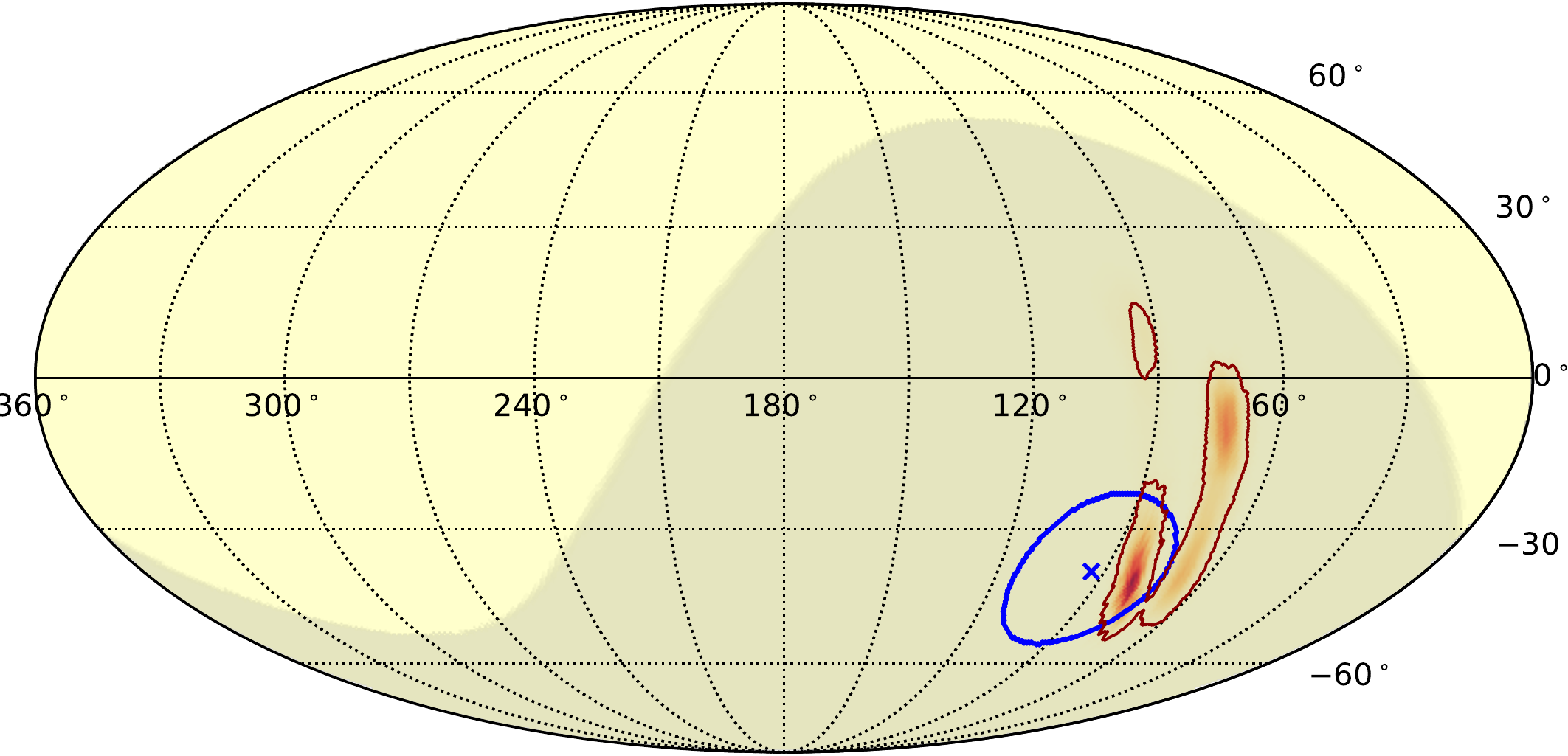}{0.36\textwidth}{(e) GW190602\_175927 (FC)}
          \fig{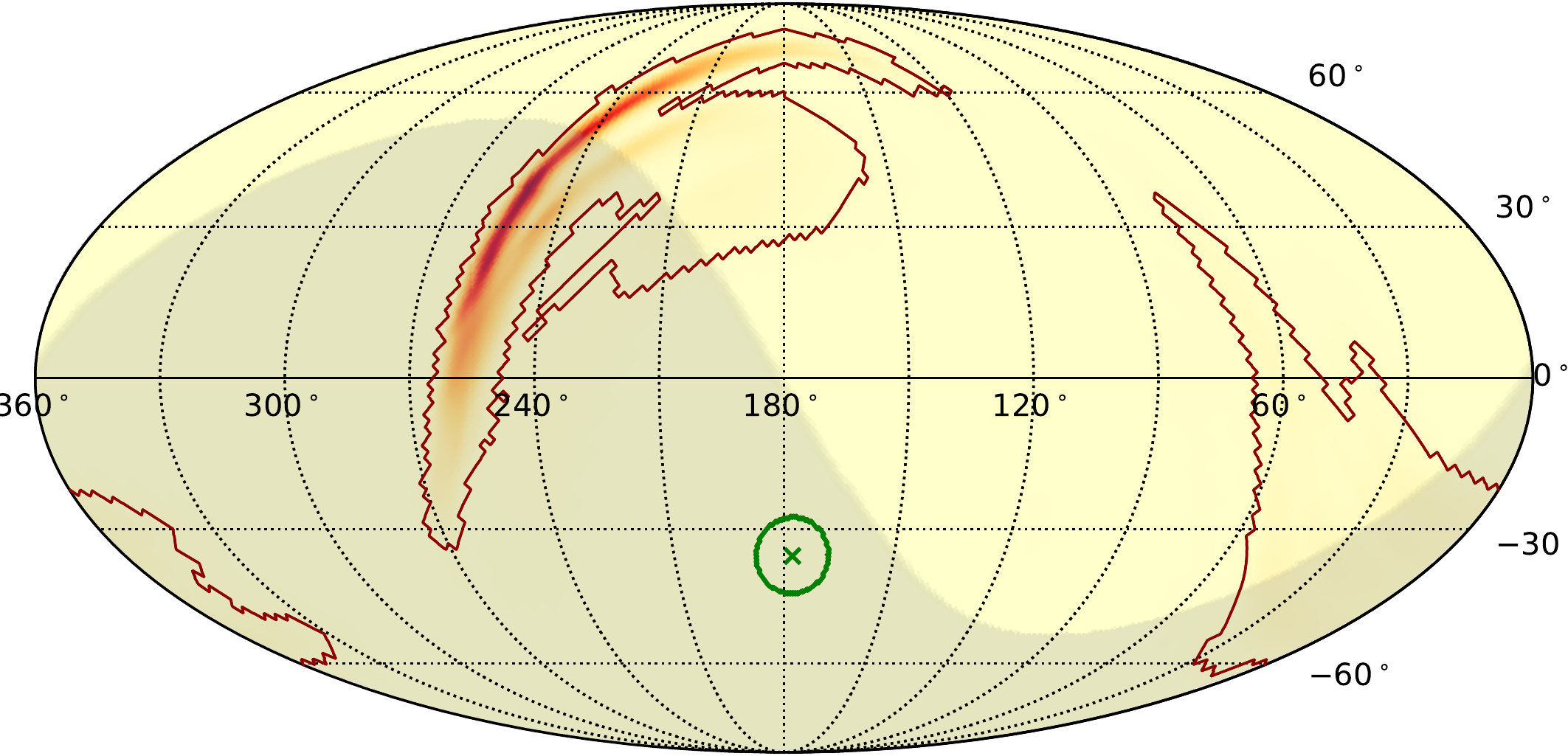}{0.36\textwidth}{(f) GW190620\_030421 (UPMU)}
}
\gridline{
          \fig{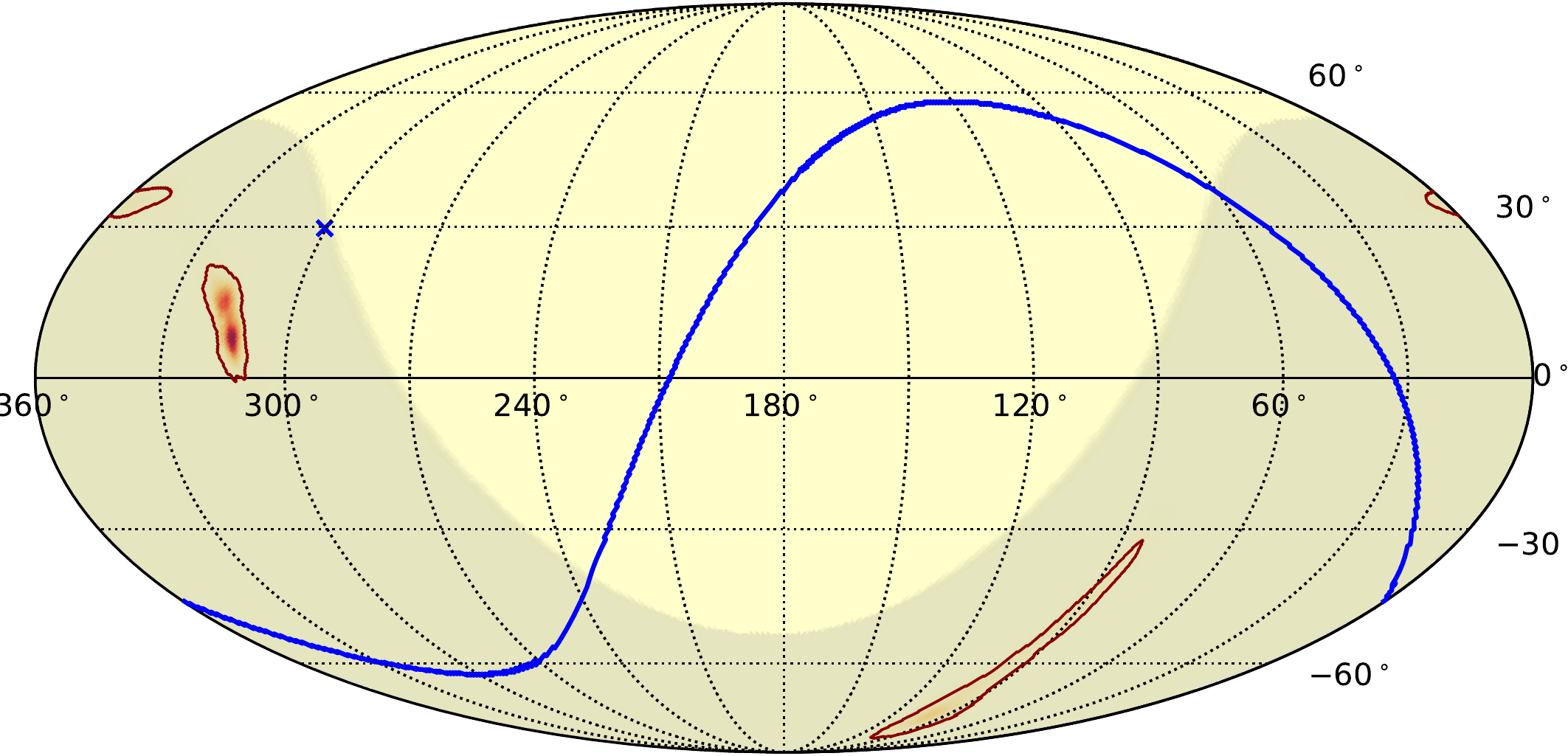}{0.36\textwidth}{(g) GW190728\_064510 (FC)}
          \fig{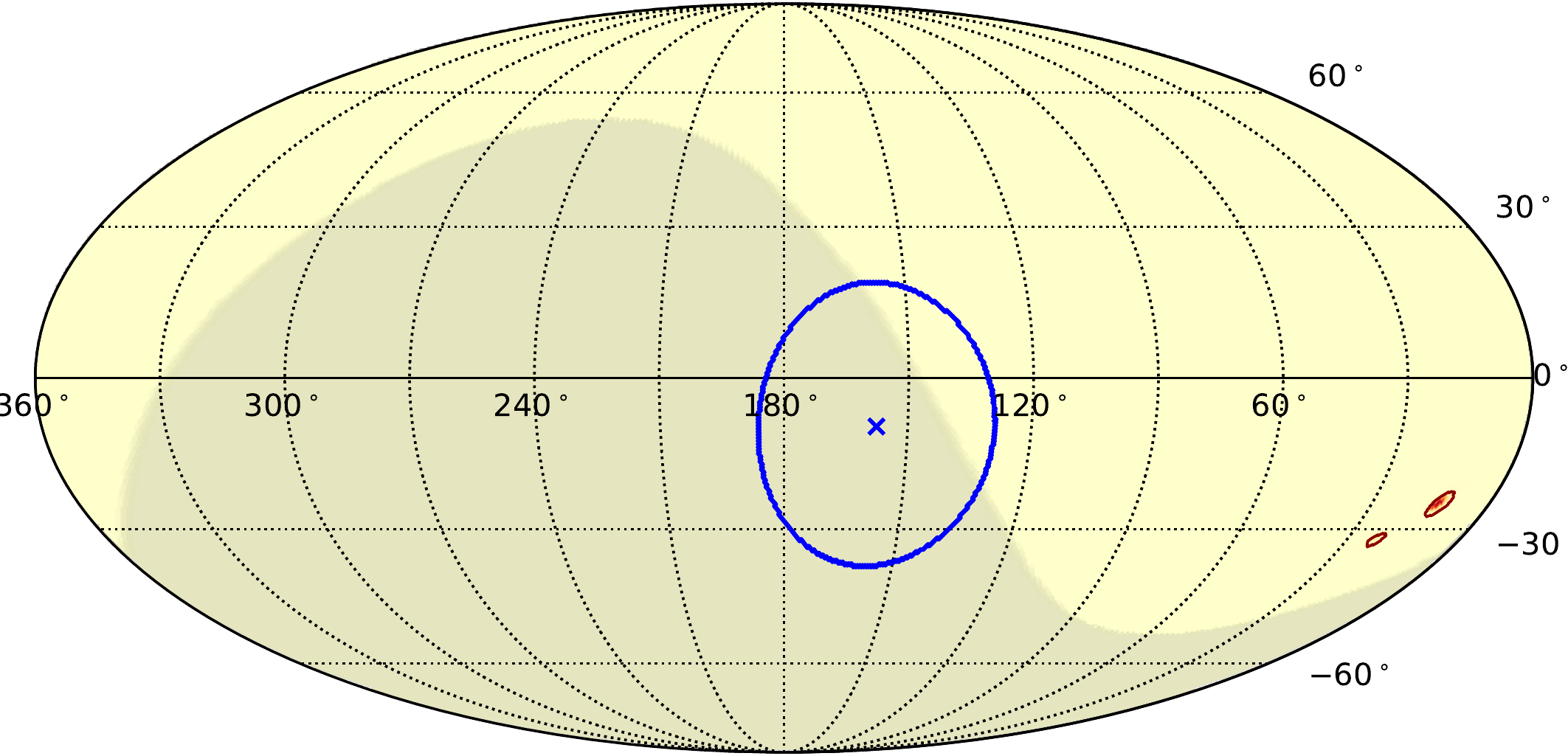}{0.36\textwidth}{(h) GW190814 (FC)}
}
\gridline{
          \fig{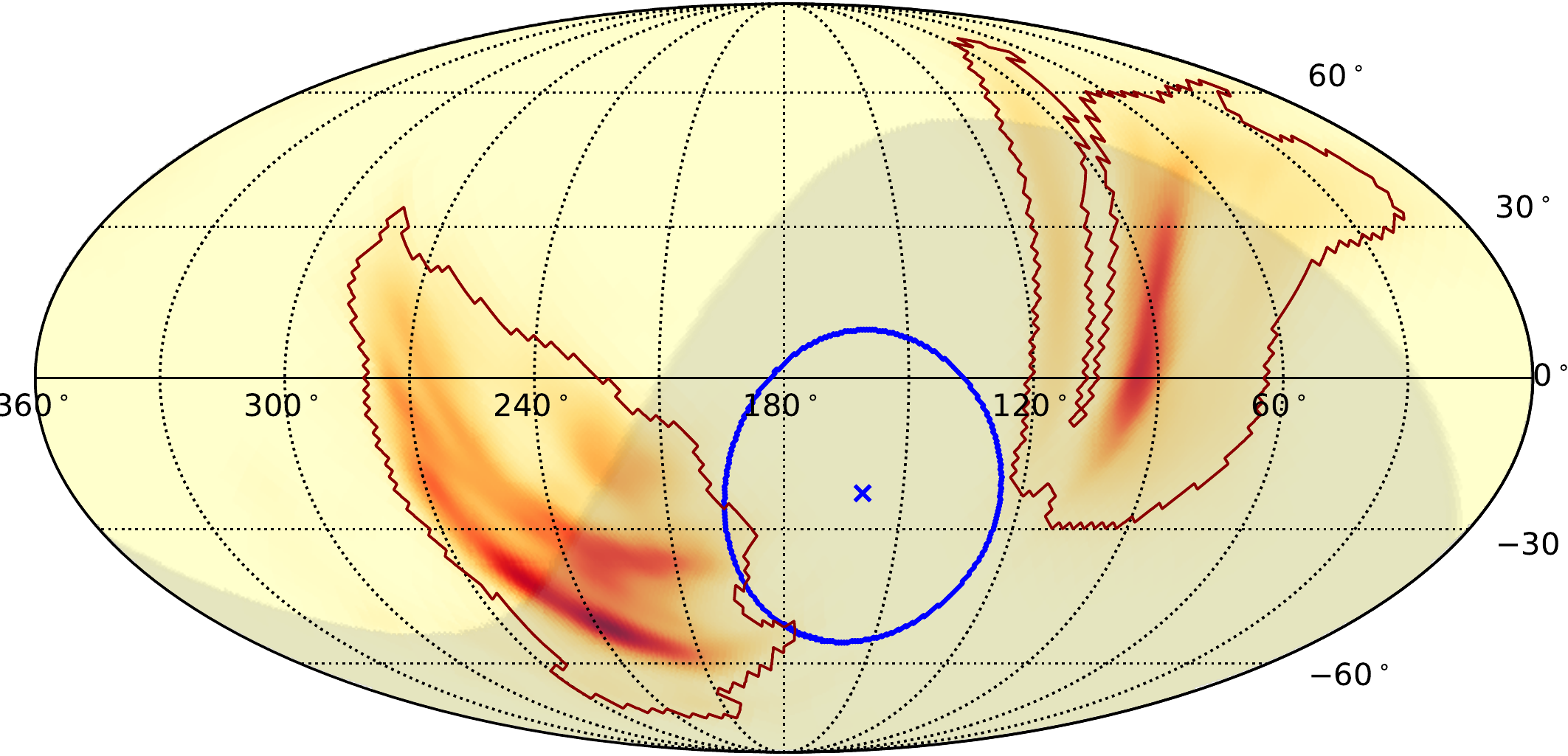}{0.36\textwidth}{(i) GW190910\_112807 (FC)}
          \fig{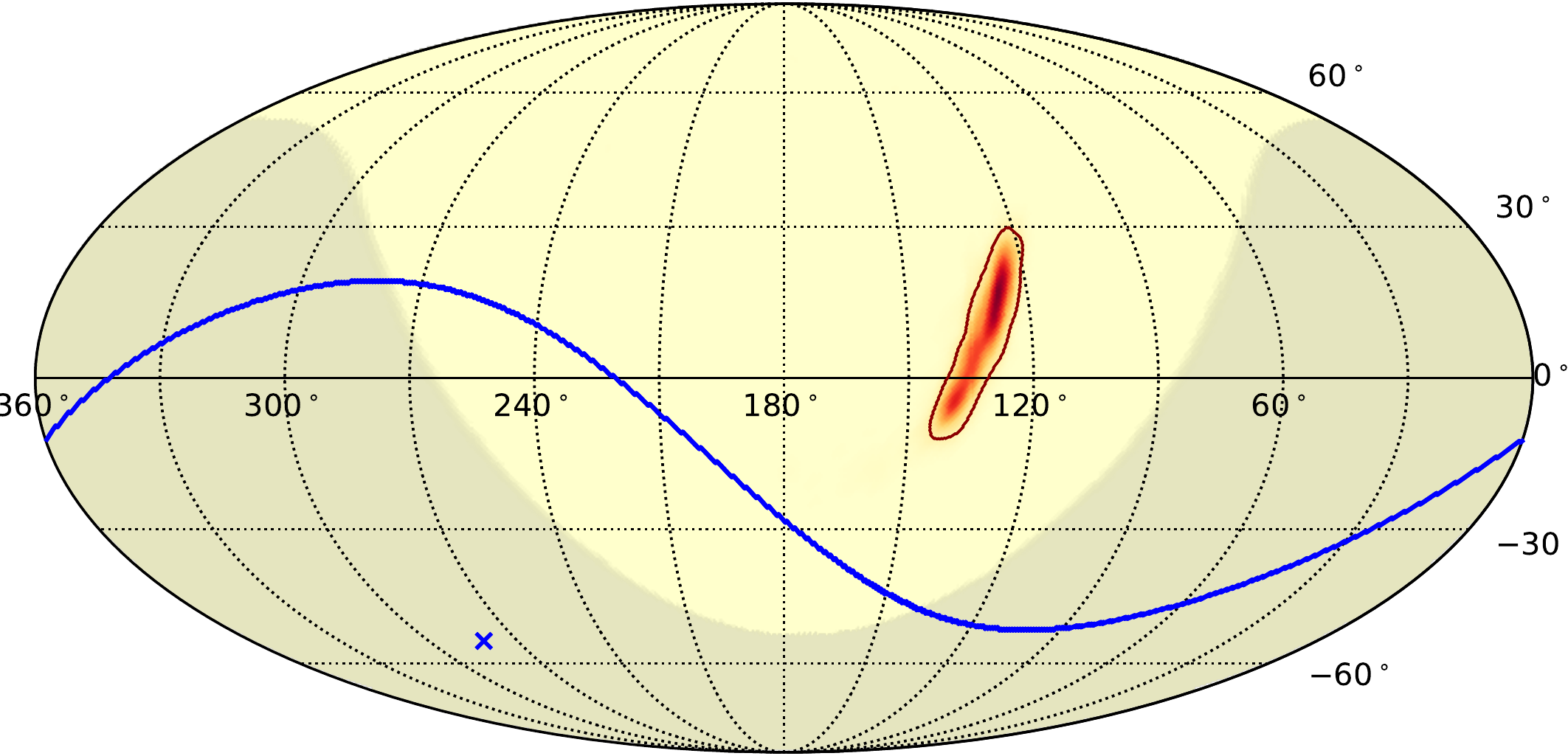}{0.36\textwidth}{(j) GW190924\_021846 (FC)}
}
          
\caption{Sky maps (in equatorial coordinates) showing the distribution of SK events, superimposed with the GW probability distribution, for the ten GW triggers with one observed event in the O3a observation run. The region representing the $1\sigma$ angular resolution is indicated in blue for FC and in green for UPMU. The dark red contour shows the 90\% containment of the GW probability. The shaded area shows the sky region that is below SK horizon (where the UPMU sample is sensitive).}
\label{fig:O3:skymaps}
\end{figure*}

The numbers of observed events in the different samples for each individual follow-up are presented in \autoref{tab:fulltable}. Out of the 36 performed follow-ups, 10 of them have associated SK HE-$\nu$ events in time coincidence (8 FC, 0 PC, 2 UPMU); this can be compared to the expected background over the 36 GWs: 4.0, 0.3, and 0.6 events, respectively, for FC, PC, and UPMU. For each selected neutrino event, the timing (in particular $\Delta t = t_{\nu} - t_{GW}$), the energy, direction, and its related angular uncertainty are provided. The latter information is presented in \autoref{tab:HEnu_table} and the angular distributions are shown in \autoref{fig:O3:skymaps}. 

\begin{deluxetable*}{l|ccc|r|cccc}
\tablecaption{Summary of all GW triggers from the O3a observation run. The first four columns summarize GW information (name, time, event type, and mean distance), the fifth column corresponds to SK live time in the \SI{1000}{\second} time window, and the last columns present the observed number of events in the four SK samples.\label{tab:fulltable}}
\tabletypesize{\small}
\tablehead{
\colhead{Trigger name} & \colhead{Alert time} & \colhead{GW Type} & \colhead{Distance} & \colhead{Live time} & \multicolumn3c{HE-$\nu$} & \colhead{LE-$\nu$} \\
\colhead{} & \colhead{(UTC)} & \colhead{} & \colhead{\si{\mega\parsec}} & \colhead{seconds} & \colhead{\footnotesize FC} & \colhead{\footnotesize PC} & \colhead{\footnotesize UPMU} & \colhead{}
}
\startdata
GW190408\_181802 & 2019-04-08 18:18:02 & BBH & 1547.5 & 993 & 0 & 0 & 0 & 3 \\ 
GW190412 & 2019-04-12 05:30:44 & BBH & 734.1 & 993 & 0 & 0 & 0 & 0 \\ 
GW190413\_052954 & 2019-04-13 05:29:54 & BBH & 4189.6 & 993 & 0 & 0 & 0 & 0 \\ 
GW190413\_134308 & 2019-04-13 13:43:08 & BBH & 5181.6 & 993 & 0 & 0 & 0 & 0 \\ 
GW190421\_213856 & 2019-04-21 21:38:56 & BBH & 3165.5 & 993 & 0 & 0 & 0 & 3 \\ 
GW190424\_180648 & 2019-04-24 18:06:48 & BBH & 2568.4 & 993 & 1 & 0 & 0 & 1 \\ 
GW190425 & 2019-04-25 08:18:05 & BNS & 156.8 & 993 & 0 & 0 & 0 & 1 \\ 
GW190426\_152155 & 2019-04-26 15:21:55 & NSBH & 377.2 & 993 & 0 & 0 & 1 & 0 \\ 
GW190503\_185404 & 2019-05-03 18:54:04 & BBH & 1527.3 & 993 & 0 & 0 & 0 & 0 \\ 
GW190512\_180714\tablenotemark{a} & 2019-05-12 18:07:14 & BBH & 1462.5 & 994 & 0 & 0 & 0 & - \\ 
GW190513\_205428 & 2019-05-13 20:54:28 & BBH & 2189.7 & 994 & 1 & 0 & 0 & 0 \\ 
GW190514\_065416 & 2019-05-14 06:54:16 & BBH & 4987.6 & 994 & 0 & 0 & 0 & 1 \\ 
GW190517\_055101\tablenotemark{b} & 2019-05-17 05:51:01 & BBH & 2270.5 & 0 & - & - & - & - \\ 
GW190519\_153544 & 2019-05-19 15:35:44 & BBH & 3023.5 & 994 & 0 & 0 & 0 & 1 \\ 
GW190521 & 2019-05-21 03:02:29 & BBH & 4566.9 & 994 & 0 & 0 & 0 & 3 \\ 
GW190521\_074359 & 2019-05-21 07:43:59 & BBH & 1244.2 & 994 & 0 & 0 & 0 & 0 \\ 
GW190527\_092055 & 2019-05-27 09:20:55 & BBH & 3562.9 & 994 & 1 & 0 & 0 & 1 \\ 
GW190602\_175927 & 2019-06-02 17:59:27 & BBH & 3138.1 & 994 & 1 & 0 & 0 & 0 \\ 
GW190620\_030421 & 2019-06-20 03:04:21 & BBH & 3210.9 & 995 & 0 & 0 & 1 & 1 \\ 
GW190630\_185205 & 2019-06-30 18:52:05 & BBH & 956.2 & 992 & 0 & 0 & 0 & 2 \\ 
GW190701\_203306 & 2019-07-01 20:33:06 & BBH & 2152.4 & 992 & 0 & 0 & 0 & 0 \\ 
GW190706\_222641 & 2019-07-06 22:26:41 & BBH & 5184.0 & 992 & 0 & 0 & 0 & 2 \\ 
GW190707\_093326 & 2019-07-07 09:33:26 & BBH & 790.8 & 992 & 0 & 0 & 0 & 0 \\ 
GW190708\_232457 & 2019-07-08 23:24:57 & BBH & 887.9 & 993 & 0 & 0 & 0 & 0 \\ 
GW190719\_215514 & 2019-07-19 21:55:14 & BBH & 4786.3 & 993 & 0 & 0 & 0 & 1 \\ 
GW190720\_000836 & 2019-07-20 00:08:36 & BBH & 906.0 & 993 & 0 & 0 & 0 & 0 \\ 
GW190727\_060333 & 2019-07-27 06:03:34 & BBH & 3608.9 & 992 & 0 & 0 & 0 & 0 \\ 
GW190728\_064510 & 2019-07-28 06:45:10 & BBH & 857.6 & 993 & 1 & 0 & 0 & 2 \\ 
GW190731\_140936 & 2019-07-31 14:09:36 & BBH & 4033.7 & 993 & 0 & 0 & 0 & 1 \\ 
GW190803\_022701 & 2019-08-03 02:27:01 & BBH & 3749.6 & 993 & 0 & 0 & 0 & 0 \\ 
GW190814 & 2019-08-14 21:10:38 & NSBH & 240.7 & 994 & 1 & 0 & 0 & 0 \\ 
GW190828\_063405 & 2019-08-28 06:34:05 & BBH & 2160.3 & 542 & 0 & 0 & 0 & 1 \\ 
GW190828\_065509\tablenotemark{b} & 2019-08-28 06:55:09 & BBH & 1657.8 & 0 & - & - & - & - \\ 
GW190909\_114149 & 2019-09-09 11:41:49 & BBH & 4923.7 & 994 & 0 & 0 & 0 & 0 \\ 
GW190910\_112807 & 2019-09-10 11:28:07 & BBH & 1670.1 & 994 & 1 & 0 & 0 & 2 \\ 
GW190915\_235702\tablenotemark{b} & 2019-09-15 23:57:02 & BBH & 1714.6 & 0 & - & - & - & - \\ 
GW190924\_021846 & 2019-09-24 02:18:46 & BBH & 572.4 & 994 & 1 & 0 & 0 & 0 \\ 
GW190929\_012149 & 2019-09-29 01:21:49 & BBH & 3901.5 & 994 & 0 & 0 & 0 & 0 \\ 
GW190930\_133541 & 2019-09-30 13:35:41 & BBH & 785.9 & 994 & 0 & 0 & 0 & 0 \\ 
\enddata
\tablenotetext{a}{The low-energy sample is not used because of high-voltage issues.}
\tablenotetext{b}{The detector was not taking data due to calibrations or maintenance.}
\end{deluxetable*}

\begin{deluxetable*}{l|c|cc|ccc|cc}
\tablecaption{List of selected SK HE-$\nu$ events in time coincidence with GW triggers from O3a observation run. The p-values obtained with the statistical method presented in \autoref{sec:pvalue} are also listed. \label{tab:HEnu_table}}
\tablewidth{0pt}
\tablehead{
    \colhead{Trigger name} & \colhead{SK sample} & \colhead{$\Delta T$} & \colhead{$E_{\rm reco}$} & \colhead{\textrm{RA}} & \colhead{\textrm{Dec}} & \colhead{$\sigma_{\rm ang}$} & \colhead{$p_{\rm space}$} & \colhead{$p$} \\
    \colhead{} & \colhead{} & \colhead{seconds} & \colhead{GeV} & \colhead{degree} & \colhead{degree} & \colhead{degree} & \colhead{\%} & \colhead{\%}
}
\startdata
GW190424\_180648 & FC & $104.03$ & $0.57$ & $210.82$ & $-58.74$ & $52.08$ & $48.55$ & $6.12$ \\ 
GW190426\_152155 & UPMU & $278.99$ & $9.52$ & $352.37$ & $-8.46$ & $2.15$ & $100.00$ & $12.59$ \\ 
GW190513\_205428 & FC & $-183.27$ & $0.68$ & $279.34$ & $-37.27$ & $41.19$ & $8.59$ & $1.08$\\ 
GW190527\_092055 & FC & $248.41$ & $0.48$ & $54.09$ & $18.80$ & $52.08$ & $58.93$ & $7.43$ \\ 
GW190602\_175927 & FC & $-286.52$ & $2.75$ & $93.67$ & $-38.90$ & $16.22$ & $1.72$ & $0.22$ \\ 
GW190620\_030421 & UPMU & $-327.70$ & $2.33$ & $177.69$ & $-35.59$ & $8.04$ & $100.00$ & $12.62$ \\ 
GW190728\_064510 & FC & $102.99$ & $0.19$ & $300.45$ & $29.71$ & $92.51$ & $21.02$ & $2.65$ \\ 
GW190814 & FC & $250.36$ & $1.21$ & $157.59$ & $-9.47$ & $28.26$ & $100.00$ & $12.61$ \\ 
GW190910\_112807 & FC & $301.42$ & $1.08$ & $160.13$ & $-22.70$ & $32.09$ & $57.11$ & $7.20$ \\ 
GW190924\_021846 & FC & $411.87$ & $0.30$ & $281.38$ & $-54.52$ & $73.58$ & $50.49$ & $6.37$ \\ 
\enddata
\end{deluxetable*}

\subsection{LE-\texorpdfstring{$\nu$}{nu} sample}

The events within the $\SI{1000}{\second}$ time window passing the selection described in \autoref{sec:sk_LE} were extracted. As in the case of HE-$\nu$ samples, the total number of observed events is compared to the background expectation. The latter is based on the average event rate computed using the total SK-V dataset, which corresponds to $0.729$ expected events in $\SI{1000}{\second}$; this background level has been found to be stable over the whole period. The results for all follow-ups are summarized in \autoref{tab:fulltable}. No significant excess was observed with respect to the expected Poisson statistics, with 24 observed events and 25.0 expected.

\section{Event-by-event statistical analysis}
\label{sec:stat}

\subsection{Observation significance}
\label{sec:pvalue}

The significance of a given observation in HE-$\nu$ samples can be quantified in terms of p-value. The latter can be divided into a temporal component $p_{\rm time}$ that is evaluating the probability to observe at least one SK event in time coincidence with the GW, and a spatial component $p_{\rm space}$ comparing the direction of reconstructed neutrinos with the GW localization: $p = p_{\rm time} \times p_{\rm space}$. This discrimination allows separating the discrete time component (due to the low expected background) from the continuous spatial component.

The term $p_{\rm time}$ is simply the Poisson probability to observe at least one event in the selected time window: $p_{\rm time} = p(N \geq 1) = 1 - e^{-n_B}$. We have $p_{\rm time} \simeq 12.6\%$ for $n_B \simeq 0.13$ (total number of expected events in \SI{1000}{\second}). The term $p_{\rm space}$ is obtained using a maximum likelihood method with the GW localization used as a spatial prior. The best-fit sky position of the potential joint source is obtained by maximizing the log-likelihood ratio, and the obtained test statistic value is compared to its expected distribution from background events to extract $p_{\rm space}$ as the probability for the observation to be compatible with the background-only hypothesis given that at least one SK event in time coincidence has been observed. The method presented in \cite{Aartsen:2020mla,Hussain:2019xzb} has been adapted to SK.

For each sample ($k =$ FC, PC or UPMU), the point-source likelihood $\mathcal{L}^{(k)}_{\nu}(n_S^{(k)}, \gamma; \Omega_S)$ is defined as:
\begin{equation}
     \mathcal{L}^{(k)}_{\nu}(n_S^{(k)}, \gamma; \Omega_S) = \frac{ e^{-(n_S^{(k)}+n_B^{(k)})} (n_S^{(k)}+n_B^{(k)})^{N^{(k)}} }{N^{(k)}!} \prod_{i=1}^{N^{(k)}} \frac{n_S^{(k)} \mathcal{S}^{(k)}(\vec{x_i}, E_i; \Omega_S, \gamma) + n_B^{(k)} \mathcal{B}^{(k)}(\vec{x_i}, E_i)}{n_S^{(k)} + n_B^{(k)}},
\label{eq:lkl}
\end{equation}
where $n_S^{(k)}$ is the number of signal events in the sample (to be fitted), $\gamma$ is the spectral index of the signal neutrino spectrum (${\rm d}n/{\rm d}E_\nu \propto E^{-\gamma}$, to be fitted as well), $\Omega_S$ is the probed source direction, $n_B^{(k)}$ is the expected number of background events in the time window, and $N^{(k)}$ is the observed number of events. $\mathcal{S}^{(k)}(\vec{x_i}, E_i; \Omega_S, \gamma)$ is the signal probability density function (pdf), which depends on reconstructed event direction $\vec{x_i}$, reconstructed event energy $E_i$, source spectral index, and direction. $\mathcal{B}^{(k)}(\vec{x_i}, E_i)$ is the background pdf, which depends solely on event information.

The $\mathcal{S}^{(k)}$ and $\mathcal{B}^{(k)}$ functions were computed and tuned for $k=$ FC, PC, UPMU independently, using atmospheric neutrino Monte Carlo simulation datasets. They are both written as the product of an angular and an energy component:
\begin{align}
    \mathcal{S}^{(k)}(\vec{x_i}, E_i; \Omega_S, \gamma) &= \mathcal{A}_S(\vec{x_i}; E_i, \Omega_S) \mathcal{E}_S(E_i; \gamma) \\
    \text{and } \mathcal{B}^{(k)}(\vec{x_i}, E_i) &= \mathcal{A}_B(\vec{x_i}) \mathcal{E}_B(E_i),
\end{align}
where the point-spread function $\mathcal{A}_S(\vec{x_i}; E_i, \Omega_S)$ is characterizing the angular resolution of the detector at the considered energy $E_i$ (it is maximal for $\vec{x_i} = \Omega_S$, i.e. neutrino in the direction of the probed point source), and the energy part $\mathcal{E}_S(E_i; \gamma)$ is the convolution of the energy spectrum ${\rm d}n/{\rm d}E_\nu \propto E^{-\gamma}$ with the energy response function $f(E_i; E_\nu)$. The functions $\mathcal{A}_B(\vec{x_i})$ and $\mathcal{E}_B(E_i)$ characterize the expected background distribution with direction and energy.

For each sample $k$ and direction $\Omega_S$, the values $\widehat{n_S^{(k)}}$ and $\widehat{\gamma^{(k)}}$ maximizing the likelihood $\mathcal{L}^{(k)}_{\nu}(n_S^{(k)}, \gamma; \Omega_S)$ were obtained using \texttt{iminuit}~\citep{Dembinski:2020}\footnote{In the implementation, as the maximization is performed independently for each sample k, the $\widehat{\gamma^{(k)}}$ may differ, even though from the physical point of view, there should be only one common value. In practice, this has almost no impact as, in most of the cases, only zero or one event are observed in SK in the time window, so that \autoref{eq:Lambda} is much simpler and only one sample contributes.}. The log-likelihood ratio $\Lambda(\Omega_S)$ was then computed:
\begin{align}
     \Lambda(\Omega_S) &= 2 \sum_k \ln \left[ \dfrac{\mathcal{L}_{\nu}(\widehat{n_S^{(k)}}, \widehat{\gamma^{(k)}}; \Omega_S)}{\mathcal{L}_{\nu}(n_S^{(k)}=0; \Omega_S)} \right] + 2 \ln \mathcal{P}_{GW}(\Omega_S),
\label{eq:Lambda}
\end{align}
where $\sum_k$ sums over the three considered samples and $\mathcal{P}_{GW}(\Omega_S)$ is the spatial prior given directly by the GW sky map.

The test statistic $TS$ was defined by finding the direction in the sky maximizing $\Lambda(\Omega)$ while scanning the full sky, after it has been divided into equal-area pixels using \texttt{HEALPix} method~\citep{Gorski:2004by} (same pixelization method as used by LVC for GW sky maps):
\begin{equation}
    TS = \underset{\Omega}{\max} \left[ \Lambda(\Omega) \right].
\label{eq:TS}
\end{equation}

Finally, this number can be used to compute $p_{\rm space}$. First, the observation in SK was used to compute $TS_{\rm data}$. Over $10^5$ background toys were generated with neutrino events distributed according to the background distribution (empty toys with zero events are not considered). For each toy, $TS$ was computed and the probability distribution function $\mathcal{P}_{\rm bkg}(TS)$ was obtained and compared to the data value:
\begin{equation}
    p_{\rm space} = \int_{TS_{\rm data}}^{\infty} \mathcal{P}_{\rm bkg}(TS)\,{\rm d}TS.
\label{eq:pvalue}
\end{equation}
The value $p_{\rm space}$ is the probability for the observation to be compatible with the background-only hypothesis given that at least one SK event in time coincidence has been observed.

\autoref{tab:HEnu_table} presents the obtained $p_{\rm space}$ for the GW triggers with at least one SK event (for the other triggers, we trivially have $p = p_{\rm space} = 100\%$). No significant deviations from the background hypothesis (uniform distribution) were observed. The most significant coincidence is associated with GW190602\_175927, with a p-value $p^{\rm best}_{\rm space} = 1.72\%$ ($p^{\rm best} = 0.22\%$), corresponding to $2.1\sigma$ (respectively $2.9\sigma$). However, one needs to take into account the total number of trials involved in the catalog search ($N=10$ for $p_{\rm space}$ as the analysis is restricted to GW with at least one SK event in time coincidence, or $N=36$ for $p$). The trial factor correction is computed by performing $10^5$ background-only pseudo-experiments and checking how often one gets $\min\{p_i\}_{i=1 \ldots N} < p^{\rm best}$. This gives post-trial values $P^{\rm best}_{\rm space} = 15.9\%$ ($1.0\sigma$) and $P^{\rm best} = 7.8\%$ ($1.4\sigma$), which are fully consistent with the background-only hypothesis.

\subsection{Flux limits using high-energy samples}
\label{sec:fluxHE}

Because no statistically significant event excess was observed within the \SI{1000}{\second} time window in the HE-$\nu$ samples, the observation can be converted to an upper limit on the incoming neutrino flux. In the first approach, this was done separately for FC, PC and UPMU samples, using a similar procedure to that in \cite{Abe:2018mic}, while a second approach used the test statistic defined in \autoref{sec:pvalue} to combine the samples.

In the following, the neutrino energy spectrum is assumed to follow a simple power law with a spectral index $\gamma = 2$, that is commonly used in such astrophysical searches (e.g. \cite{Abe:2018mic}). The neutrino flux can then be written as:
\begin{equation}
\dfrac{{\rm d}n}{{\rm d}E_\nu} = \phi_0 E_\nu^{-2}. 
\end{equation}
In the following, we will report the upper limits on $\phi_0 = E_\nu^2 \, {\rm d}n/{\rm d}E_\nu$ [in \si{\giga\electronvolt\per\square\centi\meter}], for the different samples and neutrino flavors ($\nu_\mu$, $\bar\nu_\mu$, $\nu_e$, $\bar\nu_e$).

\subsubsection{Sample-by-sample approach}
\label{sec:fluxHE:simple}

\begin{figure*}[hbtp]
\gridline{\fig{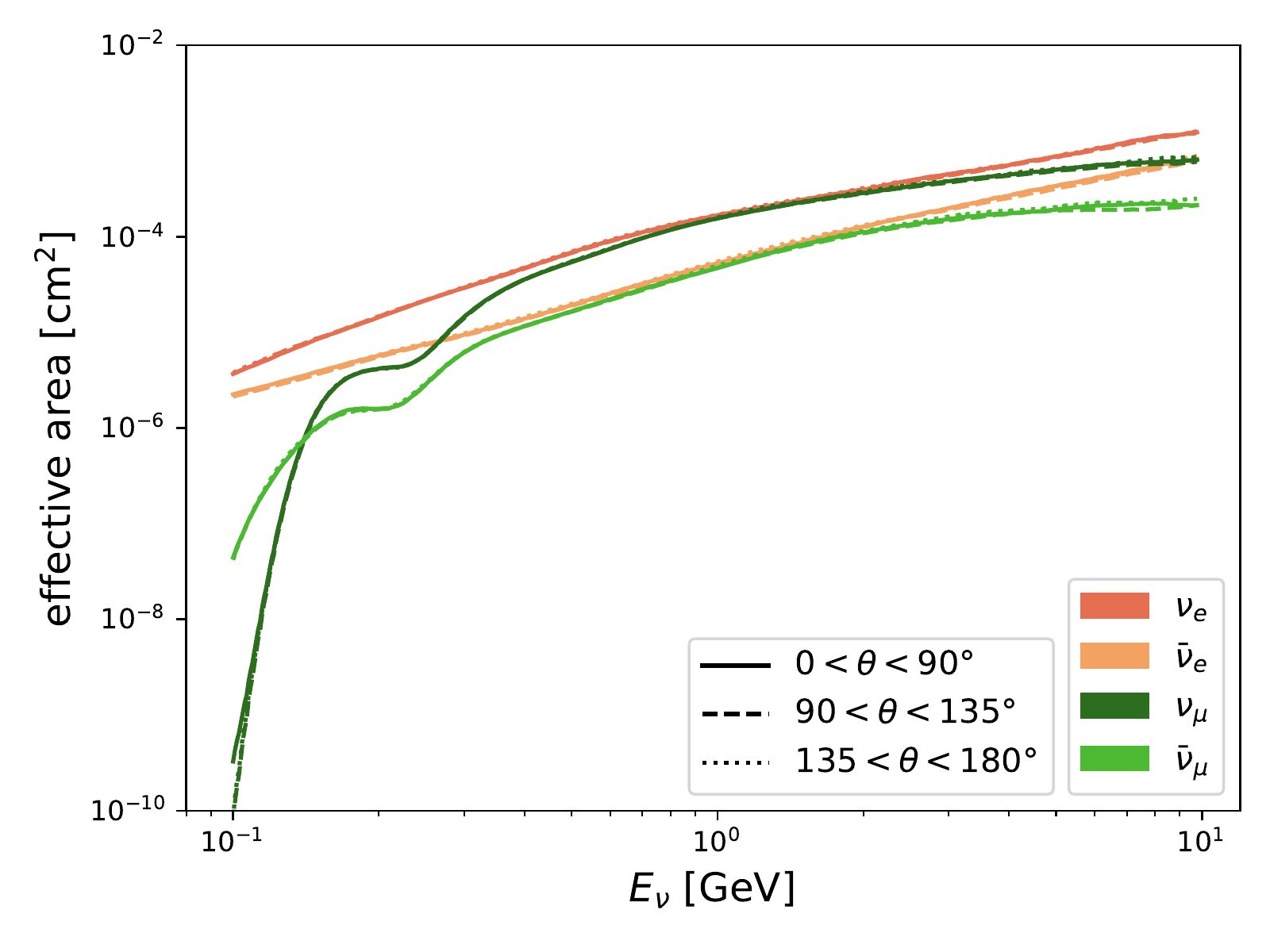}{0.48\textwidth}{(a) FC}
          \fig{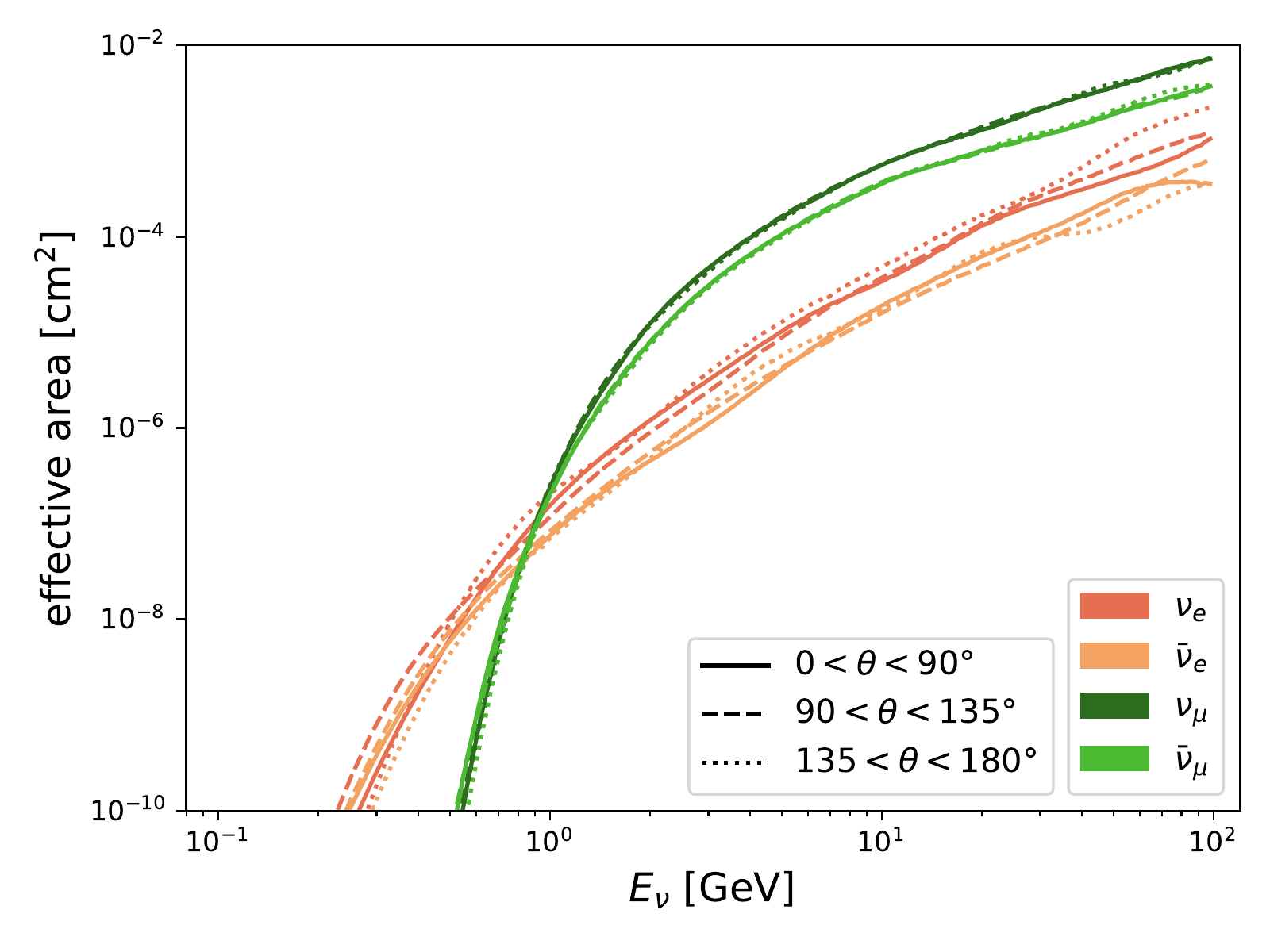}{0.48\textwidth}{(b) PC}
}
\gridline{
          \fig{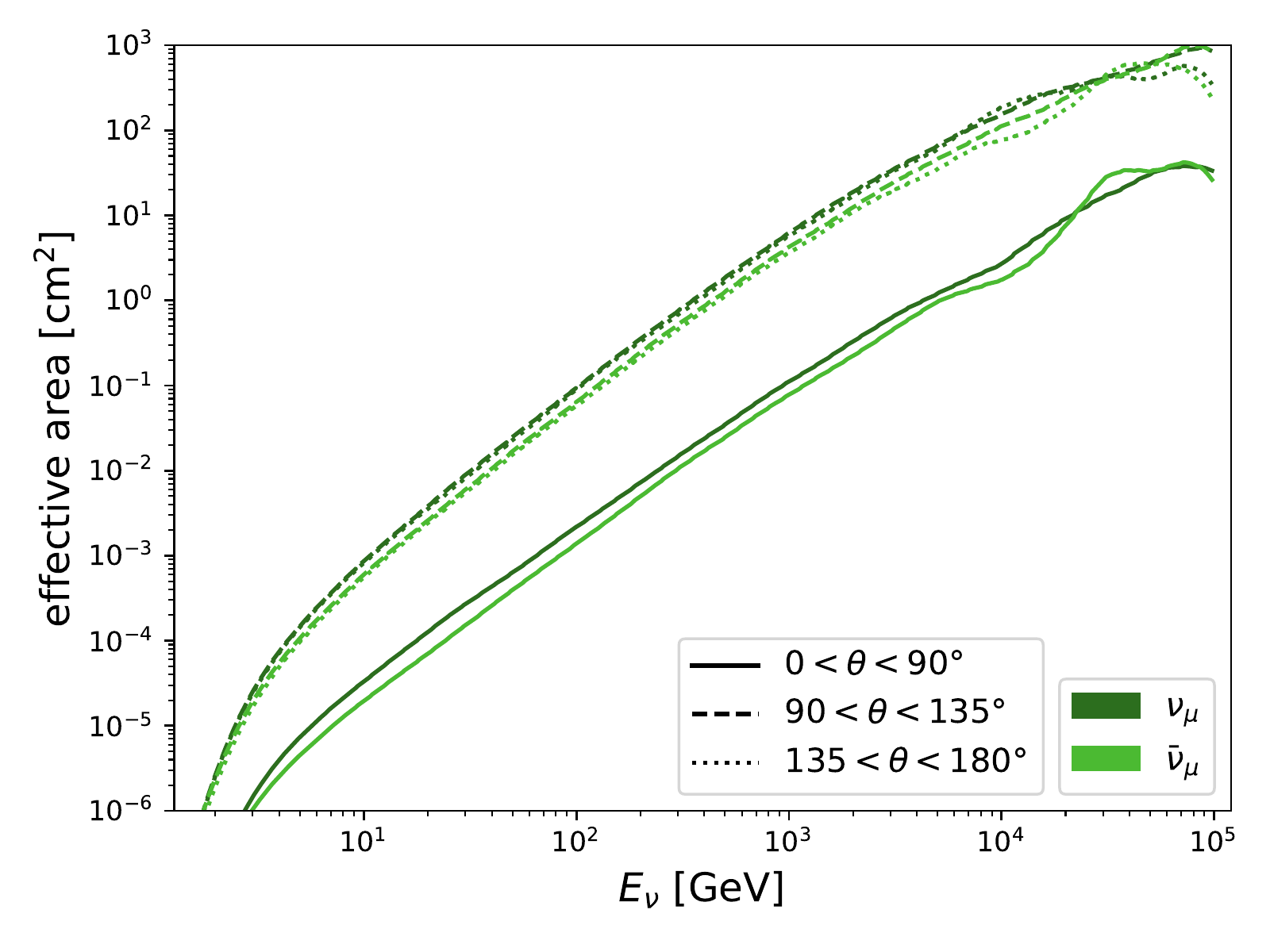}{0.48\textwidth}{(c) UPMU}
}
          
\caption{Effective area of the SK event selection, for the different HE-$\nu$ samples and neutrino flavors, as a function of neutrino energy: dark (light) green for muon \mbox{(anti-)neutrinos} and dark (light) orange for electron \mbox{(anti-)neutrinos}. The different line styles show the variation of effective area for different ranges in zenith angle.}
\label{fig:effarea}
\end{figure*}

For a given sample $s$, flavor $f$ and source position $\Omega$, the neutrino flux $E^2 {\rm d}n/{\rm d}E$ is related to the number of events:
\begin{align}
    N_{\rm sig} &= \phi_0^{(s,f)} \times \int A^{(s,f)}_{\rm eff}(E_\nu, \Omega) \times E_\nu^{-2} \, {\rm d}E_\nu \nonumber \\
                &= \phi_0^{(s,f)} \times c^{(s,f)}(\Omega),
    \label{eq:fluxconv}
\end{align}
where $A^{(s,f)}_{\rm eff}(E_\nu, \Omega)$ is the SK detector effective area for the selected sample and neutrino flavor, the integration range is $\SIrange{0.1}{e5}{\giga\electronvolt}$. The quantity $c^{(s,f)}(\Omega)$ is the detector acceptance, which takes into account the source direction. To marginalize over the source localization, the following Poisson likelihood is then defined:
\begin{equation}
    \mathcal{L}^{(s,f)}(\phi_0; n_B^{(k)}, N^{(k)}) = \displaystyle\int \dfrac{(c^{(s,f)}(\Omega) \times \phi_0+n_B^{(k)})^{N^{(k)}}}{N^{(k)}!} e^{-(c^{(s,f)}(\Omega) \times \phi_0 + n_B^{(k)})} \times \mathcal{P}_{\rm GW}(\Omega) \, {\rm d}\Omega,
\end{equation}
where $n_B^{(k)}$ and $N^{(k)}$ are respectively the expected and observed number of events in sample $s$. One can then derive 90\% confidence level (C.L.) upper limits by computing the likelihood as a function of $\phi_0$ and finding the 90\% percentile, for each sample and flavor (this effectively corresponds to the Bayesian limit with flat prior on $\phi_0$):
\begin{equation}
    \dfrac{\int_0^{\phi_0^{\rm 90\%}} \mathcal{L}^{(s,f)}(\phi_0; n_B^{(k)}, N^{(k)}) \, {\rm d}\phi_0}{\int_0^\infty \mathcal{L}^{(s,f)}(\phi_0; n_B^{(k)}, N^{(k)}) \, {\rm d}\phi_0} = 0.90.
    \label{eq:lkl_to_limit}
\end{equation}
The effective areas have been computed explicitly as a function of neutrino energy and zenith angle, using 500 years of atmospheric Monte Carlo simulations. There is a very small dependency on the local zenith angle $\theta$ for FC and PC, while UPMU covers only efficiently below the horizon ($\theta > \SI{90}{\degree}$), with a nonnegligible variation with $\theta$, as shown in \autoref{fig:effarea}. The UPMU sample has very limited sensitivity at and above the horizon ($0 < \theta < \SI{90}{\degree}$), only near-horizontal neutrinos with slightly upgoing muons can be identified. No systematic uncertainties are applied to the detector effective area estimate, as the detector response is relatively stable and well understood and the analysis is strongly dominated by limited statistics.

\begin{figure}
    \centering
    \includegraphics[width=\linewidth]{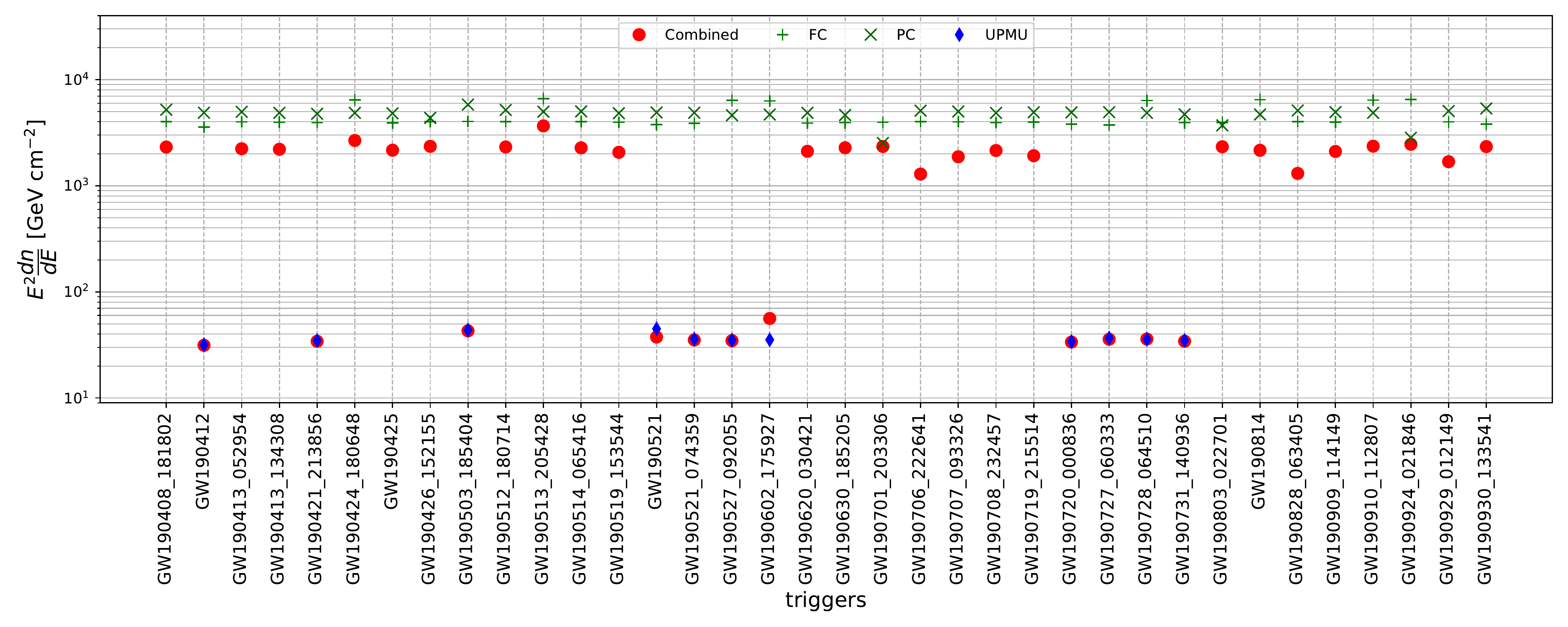}
    \caption{Obtained 90\% C.L. upper limits on $E^2 {\rm d}n/{\rm d}E$ for $\nu_\mu$ and for the different GW triggers, using the methods presented in \autoref{sec:fluxHE:simple} (sample-by-sample) and \autoref{sec:fluxHE:TS} for the combined analysis. The complete figure set (4 images, one per considered neutrino flavor) is available in the online journal.}
    \label{fig:flux_numu}
\end{figure}

The full results for $\nu_\mu$ are presented in \autoref{fig:flux_numu}. They show a wide variety of limits: in particular, UPMU upper limits are only reported if the GW localization is mainly below the horizon, where this sample has sensitivity. Detailed numbers for GW190425~\citep{Abbott:2020uma} and GW190521~\citep{Abbott:2020tfl} are presented in \autoref{tab:flux}. These two events are illustrating the two scenarios and are the only BNS candidate in GWTC-2 and the heaviest BBH, respectively. Results for all the triggers are given in \autoref{tab:datarelease}.

\begin{deluxetable*}{c|cc|cccc}
\tablecaption{Obtained 90\% C.L. upper limits on $E^2 {\rm d}n/{\rm d}E$ for GW190425 and GW190521.
For HE-$\nu$, limits on $E^2 {\rm d}n/{\rm d}E$ [in \si{\giga\electronvolt\per\square\centi\meter}] are presented for the different neutrino flavors, assuming $E^{-2}$ spectrum. Upper limits on the total energy emitted by the source as neutrinos $E_{\rm iso}$ [in erg] (assuming isotropic emission) are also presented: one limit per flavor and limits for $\nu_e$ + $\bar\nu_e$, $\nu_\mu$ + $\bar\nu_\mu$ and on the total energy in all flavors assuming equipartition (including unseen tau neutrinos). For LE-$\nu$, limits on the total neutrino fluence $\Phi$ [in \si{\per\square\centi\meter}] are given for $\nu_e$, $\bar\nu_e$, $\nu_x=\nu_\mu+\nu_\tau$, $\bar\nu_x=\bar\nu_\mu+\bar\nu_\tau$ assuming Fermi-Dirac spectrum (with average energy of \SI{20}{\mega\electronvolt}) and flat spectrum (within the range \SIrange{7}{100}{\mega\electronvolt}), as well as upper limits on $E_{\rm iso}$ [in erg] for the Fermi-Dirac scenario. \label{tab:flux}}
\tablewidth{0pt}
\tablehead{
\colhead{Trigger name} & \multicolumn2c{Sample} & \colhead{$\nu_e$} & \colhead{$\bar\nu_e$} & \colhead{$\nu_\mu$ ($\nu_x$)} & \colhead{$\bar\nu_\mu$ ($\bar\nu_x$)}
}
\startdata
 & HE $E^2 \dfrac{{\rm d}n}{{\rm d}E}$ & FC & $2.22 \cdot 10^{3}$ & $4.32 \cdot 10^{3}$ & $3.91 \cdot 10^{3}$ & $9.42 \cdot 10^{3}$ \\ 
 &    & PC & $3.32 \cdot 10^{4}$ & $1.12 \cdot 10^{5}$ & $4.81 \cdot 10^{3}$ & $8.74 \cdot 10^{3}$ \\ 
 &    & UPMU & $-$ & $-$ & $-$ & $-$ \\ 
 &    & Combined & $2.09 \cdot 10^{3}$ & $4.28 \cdot 10^{3}$ & $2.16 \cdot 10^{3}$ & $4.20 \cdot 10^{3}$ \\ 
\cline{2-7} 
 & HE $E_{\rm iso}$ & Per-flavour & $1.98 \cdot 10^{56}$ & $3.85 \cdot 10^{56}$ & $1.96 \cdot 10^{56}$ & $3.69 \cdot 10^{56}$ \\ 
GW190425 &    & $\nu + \bar\nu$ & \multicolumn{2}{c}{$2.62 \cdot 10^{56}$} & \multicolumn{2}{c}{$2.52 \cdot 10^{56}$} \\ 
 &    & All & \multicolumn{4}{c}{$3.47 \cdot 10^{56}$} \\ 
\cline{2-7} 
 & LE $\Phi$ & Flat & $1.49 \cdot 10^{9}$ & $1.83 \cdot 10^{7}$ & $9.35 \cdot 10^{9}$ & $1.11 \cdot 10^{10}$ \\ 
 &    & Fermi-Dirac & $3.92 \cdot 10^{9}$ & $9.57 \cdot 10^{7}$ & $2.43 \cdot 10^{10}$ & $2.87 \cdot 10^{10}$ \\ 
\cline{2-7} 
 & LE $E_{\rm iso}$ & Per-flavour & $3.92 \cdot 10^{59}$ & $9.59 \cdot 10^{57}$ & $2.43 \cdot 10^{60}$ & $2.87 \cdot 10^{60}$ \\ 
 &    & All & \multicolumn{4}{c}{$5.54 \cdot 10^{58}$} \\ 
\hline 
 & HE $E^2 \dfrac{{\rm d}n}{{\rm d}E}$ & FC & $2.27 \cdot 10^{3}$ & $4.71 \cdot 10^{3}$ & $3.76 \cdot 10^{3}$ & $9.60 \cdot 10^{3}$ \\ 
 &    & PC & $3.66 \cdot 10^{4}$ & $3.68 \cdot 10^{4}$ & $4.89 \cdot 10^{3}$ & $8.35 \cdot 10^{3}$ \\ 
 &    & UPMU & $-$ & $-$ & $4.48 \cdot 10^{1}$ & $5.04 \cdot 10^{1}$ \\ 
 &    & Combined & $2.21 \cdot 10^{3}$ & $4.60 \cdot 10^{3}$ & $3.75 \cdot 10^{1}$ & $4.82 \cdot 10^{1}$ \\ 
\cline{2-7} 
 & HE $E_{\rm iso}$ & Per-flavour & $1.69 \cdot 10^{59}$ & $3.46 \cdot 10^{59}$ & $2.58 \cdot 10^{57}$ & $3.72 \cdot 10^{57}$ \\ 
GW190521 &    & $\nu + \bar\nu$ & \multicolumn{2}{c}{$2.26 \cdot 10^{59}$} & \multicolumn{2}{c}{$3.00 \cdot 10^{57}$} \\ 
 &    & All & \multicolumn{4}{c}{$8.94 \cdot 10^{57}$} \\ 
\cline{2-7} 
 & LE $\Phi$ & Flat & $2.63 \cdot 10^{9}$ & $3.22 \cdot 10^{7}$ & $1.65 \cdot 10^{10}$ & $1.95 \cdot 10^{10}$ \\ 
 &    & Fermi-Dirac & $6.89 \cdot 10^{9}$ & $1.68 \cdot 10^{8}$ & $4.27 \cdot 10^{10}$ & $5.04 \cdot 10^{10}$ \\ 
\cline{2-7} 
 & LE $E_{\rm iso}$ & Per-flavour & $5.85 \cdot 10^{62}$ & $1.43 \cdot 10^{61}$ & $3.63 \cdot 10^{63}$ & $4.28 \cdot 10^{63}$ \\ 
 &    & All & \multicolumn{4}{c}{$8.26 \cdot 10^{61}$} \\ 
\enddata
\end{deluxetable*}

\subsubsection{Combination of the samples (using test statistic)}
\label{sec:fluxHE:TS}

As the neutrino spectrum is expected to span the full range from \SI{0.1}{\giga\electronvolt} to \SI{e5}{\giga\electronvolt}, it is worth combining the different samples that have varying sensitivities (in energy, flavor and direction). The method initially presented in \cite{Veske:2020yjt} was implemented using the test statistic defined previously.

Signal simulations were performed, assuming $E^{-2}$ spectrum and that at most two signal neutrinos are detected in SK; the source direction is chosen randomly based on GW sky map $\mathcal{P}_{GW}$ and the distribution of signal toy events between the samples is done according to the relative effective areas. As with the background toys in \autoref{sec:pvalue}, this allows computing the pdf $\mathcal{P}_{n_S}(TS)$ for a given number of signal events $n_S = 0,1,2$.

Assuming that at most two signal neutrinos will be observed for a given GW trigger, the following flux likelihood is defined, based on the observed test statistic and GW sky map:
\begin{equation}
    \mathcal{L}(\phi_0; TS_{\rm data}, \mathcal{P}_{GW}) = \int 
        \sum_{k=0}^{2} \Big[ \dfrac{\left( c(\Omega) \phi_0 \right)^k}{k!} e^{-c(\Omega) \phi_0} \times \mathcal{P}_k(TS_{\rm data}) \Big] \times \mathcal{P}_{GW}(\Omega) \, {\rm d}\Omega,
    \label{eq:fluxlkl}
\end{equation}
where $c(\Omega)=\sum_s c^{(s,f)}(\Omega)$ is the total detector acceptance (summing all samples) assuming $E^{-2}$ spectrum and the other quantities have already been defined above. The likelihood is composed of a sum of Poisson terms that quantify the relation between number of events and the flux, weighted by the probability to observe the measured test statistic given the different signal hypotheses.

The 90\% C.L. upper limit on $\phi_0 = E_\nu^2 {\rm d}n/{\rm d}E_\nu$ is then simply obtained as in \autoref{eq:lkl_to_limit}. The procedure can be repeated independently for each neutrino flavor or also combining flavors, e.g., $\nu_\mu + \bar\nu_\mu$ (in the latter case, both $\mathcal{P}_{n_S}(TS)$ terms and $c(\Omega)$ are computed assuming equally distributed flux between the different flavors). The results are presented in \autoref{fig:flux_numu} and \autoref{tab:flux} for the two examples mentioned above, and in \autoref{tab:datarelease} for all the events.

The combined limits are usually close to the limits obtained by the most constraining individual sample. If the UPMU sample is used (GW localized mainly below the horizon), the combined limit is similar to the UPMU limit. Otherwise, it is consistent with the result of FC+PC. In the case of GW190602\_175927, the combined limit is slightly worse than the individual UPMU because of the observed FC event in the same direction as the GW, which gives higher $TS_{\rm data}$ and thus impact $\mathcal{P}_k(TS_{\rm data})$ used in the \autoref{eq:fluxlkl}.

\subsection{Flux limits using Low-energy sample}
\label{sec:fluxLE}

The flux limit calculation for the low-energy sample is similar to HE-$\nu$, except that the effective area is parameterized as in \cite{Abe:2018mic}. As there is no direction dependence of the latter and there is only one LE-$\nu$ sample, there is no need to define a likelihood in order to perform a combination or to marginalize over the sky like in the HE-$\nu$ case.

The upper limit on the total fluence ($\Phi = \int {\rm d}n/{\rm d}E \; {\rm d}E$) is then simply computed as:
\begin{equation}
    \Phi_{90} = \frac{N_{90}}{N_{T}\int \lambda(E_{\nu}) \sigma(E_{\nu}) R(E_{e},E_{\rm vis}) \epsilon(E_{\rm vis}) \, {\rm d}E_{\nu}},
\end{equation}
where $N_{90}$ is the 90\% C.L. upper limit on the number of signal events (calculated from a Poisson distribution), $N_{T}$ is the number of target nuclei in SK fiducial volume, $\sigma$ is the combined cross section for all interactions, $\epsilon$ is the detection efficiency, $\lambda$ is the energy density assuming a given spectrum (${\rm d}n/{\rm d}E_\nu = \Phi \times \lambda(E)$) and $R$ is the response function to convert electron or positron energy ($E_{e}$) to visible energy in SK ($E_{\rm vis}$). The response function and the detection efficiency ($\epsilon$) are calculated using SK detector Monte Carlo simulations, and related systematic uncertainties are neglected as in the HE-$\nu$ case.

In this analysis, two types of spectra were considered: flat spectrum ($\lambda = \textrm{constant}$) and Fermi-Dirac spectrum with an average energy of \SI{20}{\mega\electronvolt}. The results for a selection of triggers are shown in \autoref{tab:flux}, while the rest are detailed in \autoref{tab:datarelease}. The limits are more stringent for the $\bar\nu_e$ case, given that the main interaction channel in the detector is inverse beta decay of $\bar\nu_e$, as described in \autoref{sec:sk}.

\section{Neutrino emission limits and population constraints}
\label{sec:pop}

None of the joint observations has a significance high enough in order to classify them as detection (as presented in \autoref{tab:HEnu_table}) and the flux limits provided in the previous sections do not directly constrain the physical quantities related to the astrophysical objects. In this section, the neutrino emission at the source is assumed to be isotropic, so that the intrinsic energy $E_{\rm iso}$ emitted by neutrinos from a source at a distance $d$ is directly related to the detected flux at Earth:
\begin{equation}
    E_{\rm iso} = 4\pi d^2 \int \frac{{\rm d}n}{{\rm d}E} \times E \, {\rm d}E.
    \label{eq:Eiso}
\end{equation}
Knowing $d$, one can then constrain $E_{\rm iso}$, as described in the following.

\subsection{High-energy neutrino emission}
\label{sec:popHE}

If, as in \autoref{sec:fluxHE}, $E^ {-2}$ spectrum is assumed, the \autoref{eq:Eiso} can then be integrated under this particular assumption: 
\begin{equation}
    \dfrac{E_{\rm iso}}{4\pi d^2} = \int \phi_0 E^{-2} \times E \, {\rm d}E = \phi_0 \times \ln \left( \dfrac{E_{\rm max}}{E_{\rm min}} \right).
\end{equation}
To use the GW sky map as an input, the following likelihood is defined \citep{Veske:2020yjt}:
\begin{equation}
    \mathcal{L}(E_{\rm iso}; TS_{\rm data}^{(i)}, \mathcal{V}_{GW}^{(i)}) =
    \int \sum_{k=0}^{2} \Big[ \dfrac{\left( c'(r,\Omega) E_{\rm iso} \right)^k}{k!} e^{-c'(r,\Omega) E_{\rm iso}} \times \mathcal{P}^{(i)}_k(TS^{(i)}_{\rm data}) \Big] \times \mathcal{V}^{(i)}_{GW}(r,\Omega) {\rm d}r {\rm d}\Omega.
\end{equation}
The quantity $c'(r,\Omega)$ is the conversion factor from $E_{\rm iso}$ to the expected number of signal events in SK for known source distance $r$ and direction $\Omega$. The test statistic distributions $\mathcal{P}^{(i)}_k(TS)$ and the measured test statistic $TS^{(i)}_{\rm data}$ for trigger $i$ are the same as defined in \autoref{sec:fluxHE:TS}. $\mathcal{V}_{GW}^{(i)}(r, \Omega)$ is the three-dimensional LVC sky map provided for trigger $i$, taking into account both the direction localization and the distance to the source (see \cite{Singer:2016eax} for details on GW 3D localization). 

One can derive $E_{\rm iso}$ limits independently for each GW trigger as it has been done for the flux limits. For a given flavor (e.g., $\nu_\mu$), the obtained limit is on the isotropic energy emitted from the source and that would be detected with this given flavor on Earth (with no assumptions on the flavor distribution). Limits on the total energy emitted by neutrinos of all flavors can be obtained by considering all detectable flavors in SK and assuming equal proportions of them at Earth. This is a reasonable assumption in the most common source scenario, where neutrinos are produced in pion decays in a flavor ratio $(\nu_e:\nu_\mu:\nu_\tau)$ equal to $(1:2:0)$, which would become $\sim (1:1:1)$ at Earth, after oscillations.

\begin{figure}
    \gridline{
        \fig{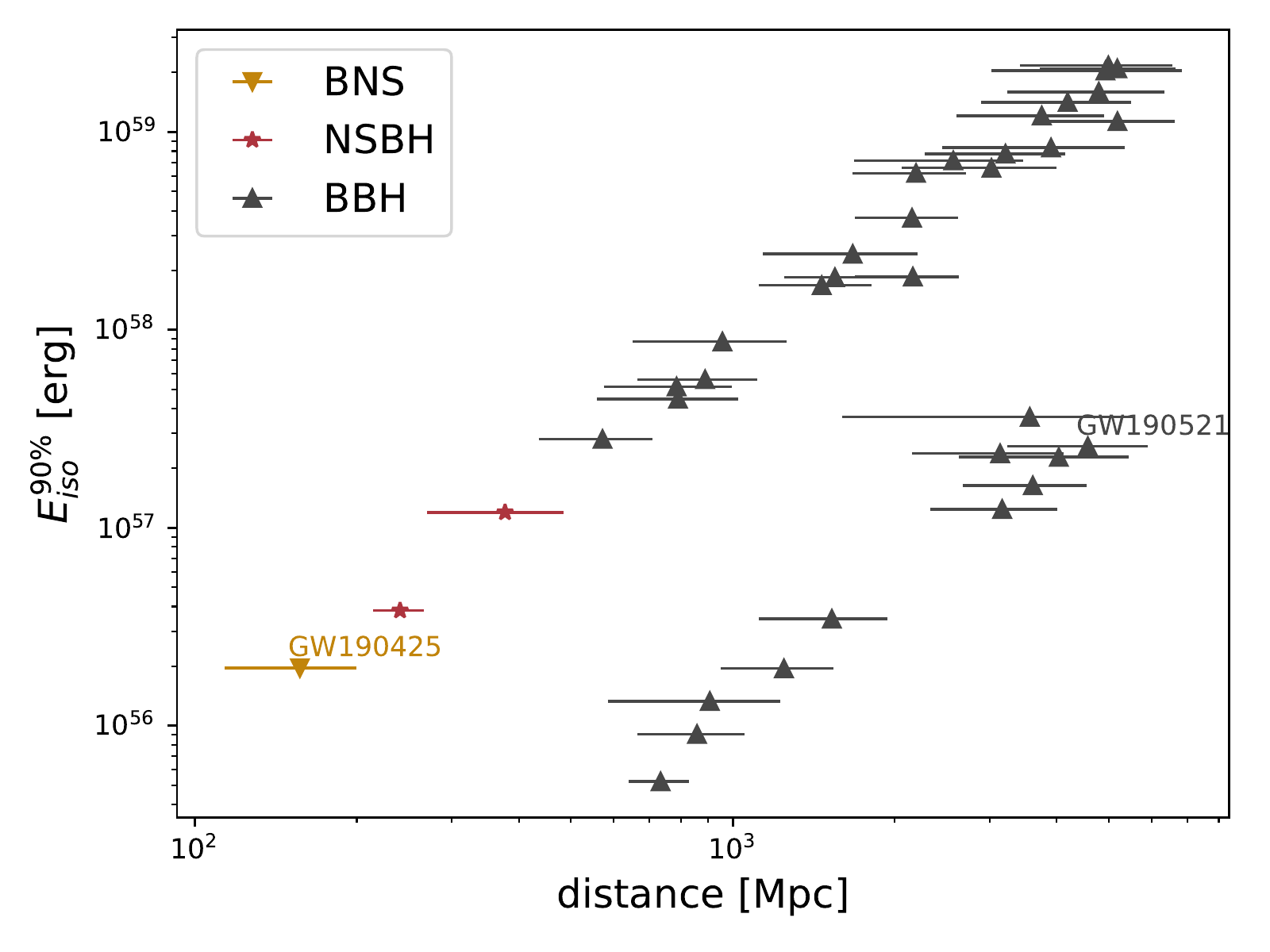}{0.48\textwidth}{(a) Limits on $E_{\rm iso}^{\nu_\mu}$}
        \fig{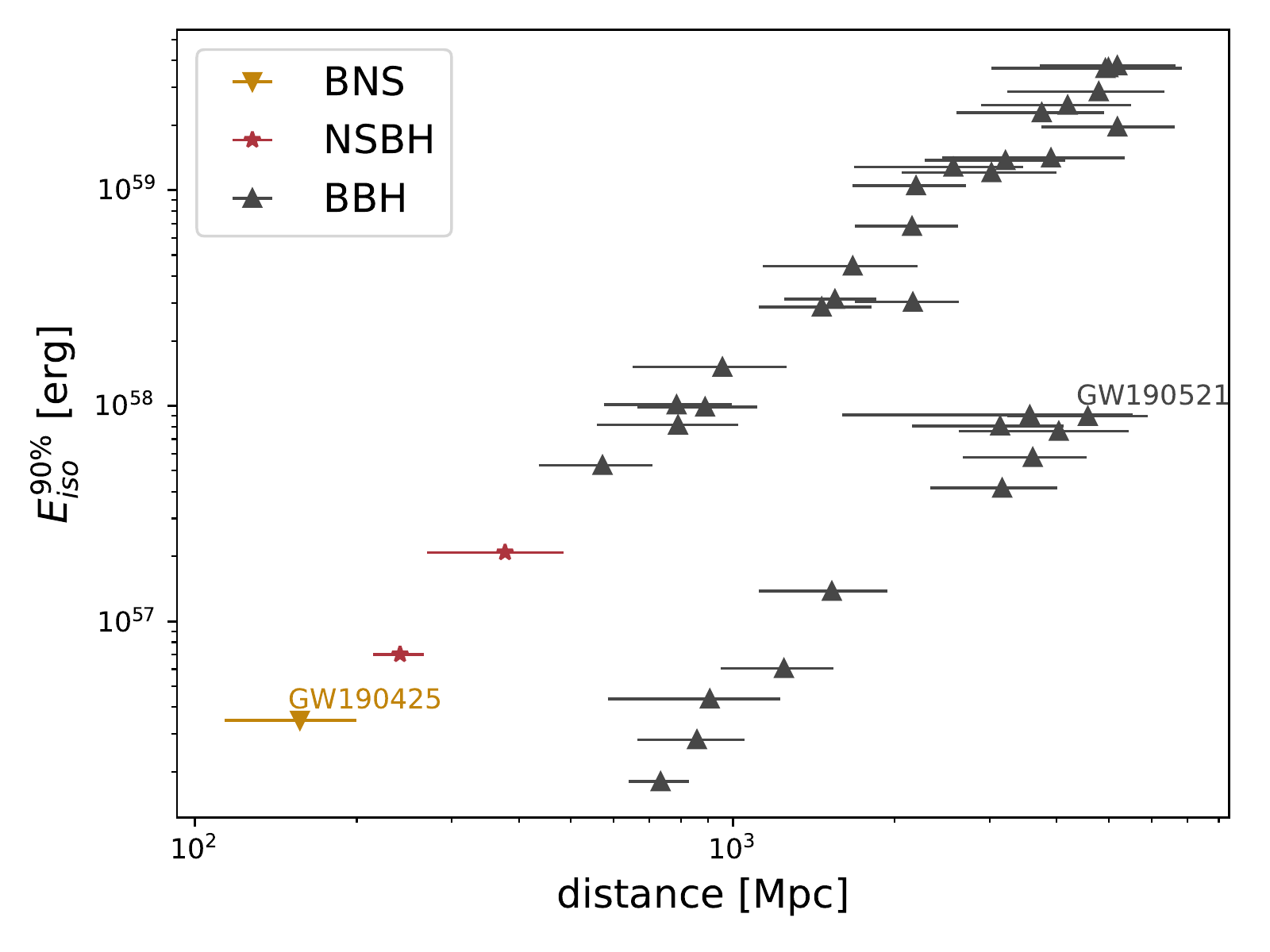}{0.48\textwidth}{(b) Limits on $E_{\rm iso}^{\rm all-flavors}$}
    }
    \caption{90\% C.L. upper limits on the isotropic energy emitted in neutrinos for the 36 GW triggers followed up by SK, as a function of source distance. The distance and its error, as well as the source type (indicated by the different colors and markers), are provided using the data from \cite{Abbott:2020niy} ($m < \SI{3}{\solarmass}$ = NS, $m > \SI{3}{\solarmass}$ = BH). The limits are following two lines $E_{\rm iso}^{90\%} \propto \text{distance}^2$ based on geometrical considerations, one of the lines shows events dominated by UPMU $\nu_\nu/\bar\nu_\mu$ contributions (giving more stringent limits) while the other line contains GW triggers that are less constrained. The two GW used in \autoref{tab:flux} are labelled in the plots. The complete figure set (5 images, one per considered neutrino flavor + all-flavors) is available in the online journal.}
    \label{fig:Eiso}
\end{figure}

The results are detailed for a selection of triggers in \autoref{tab:flux} and are plotted in \autoref{fig:Eiso}; the full results are shown in \autoref{tab:datarelease}. In the example of GW190521, the UPMU sample contributed to the observation so that the most constraining limits are obtained for $\nu_\mu$ and $\bar\nu_\mu$ ; the limit on the total energy emitted in neutrinos assuming equipartition is then dominated by the latter contributions: $E_{\rm iso,90\%}^{\rm all-flavors} \simeq 3 \times E_{\rm iso,90\%}^{\nu_\mu + \bar\nu_\mu}$. Instead, for GW190425, the limit has similar contributions from all neutrino flavors, as the UPMU sample is not contributing to the limit.

It is worth mentioning that the only BNS in the catalog, GW190425, is located in a sky region for which the observation with the UPMU sample is not possible, as already mentioned in \autoref{sec:fluxHE:TS}. If it had been located in a more favorable region, the upper limit would have improved by a factor $\sim 30$. This is promising for future observations.

Furthermore, if the spectrum is assumed to follow a $E^{-3}$ spectrum, all the limits presented above are getting less constraining, due to this less favorable spectrum (shifted to lower energies where associated effective areas are smaller), as detailed in \autoref{tab:datarelease} for the combined all-flavor $E_{\rm iso}$ limits.

The combination of a meaningful set of GW events to constrain further $E_{\rm iso}$ is also worthwhile to infer information about the common physical processes involved in a given source population. This can be performed for different sets of triggers, based on the classification provided in the GW catalog. The relevant categories are BBH, BNS, and NSBH. If emission from all objects of the same nature is assumed to be similar (independently of their individual characteristics), one can define the likelihood:
\begin{equation}
\mathcal{L}^{\rm Pop}(E_{\rm iso}; \{ TS_{\rm data})^{(i)} \}, \{ \mathcal{V}_{GW}^{(i)} \}) = \prod_{i=1}^N \mathcal{L}(E_{\rm iso}; TS_{\rm data})^{(i)}, \mathcal{V}_{GW}^{(i)}),
\end{equation}
where the sum runs over the selected GW triggers to be combined.

A more realistic toy scenario would be that the neutrino emission scales with the total mass $\mathcal{M}_{\rm tot}$ of the binary system: $E_{\rm iso}^{\nu} = f_\nu \times \mathcal{M}_{\rm tot}$. One can then use the following likelihood to constrain $f_\nu$:
\begin{equation}
\mathcal{L}^{\rm Pop}(f_\nu; \{ TS_{\rm data}^{(i)} \}, \{ \mathcal{V}_{GW}^{(i)} \}, \{ \mathcal{M}_{\rm tot}^{(i)} \}) = \prod_{i=1}^N \int \mathcal{M}_{\rm tot}^{(i)} \times \mathcal{L}(f_\nu \mathcal{M}_{\rm tot}^{(i)}; TS_{\rm data}^{(i)}, \mathcal{V}_{GW}^{(i)}) \times p_{\rm GW}(\mathcal{M}_{\rm tot}^{(i)}) \times {\rm d}\mathcal{M}_{\rm tot}^{(i)},
\end{equation}
where $f_\nu$, in \si[per-mode=symbol]{\erg\per\solarmass} is to be constrained (simplifying the units, $f_\nu$ can be expressed as the proportion of the total mass converted in neutrinos: e.g. $f_\nu = \SI[per-mode=symbol]{e54}{\erg\per\solarmass} = 62\%$), and $p_{\rm GW}(\mathcal{M}_{\rm tot}^{(i)})$ is the posterior distribution of the total mass of the binary system, as obtained from the LVC data release.

\begin{figure}
    \centering
    \gridline{
        \fig{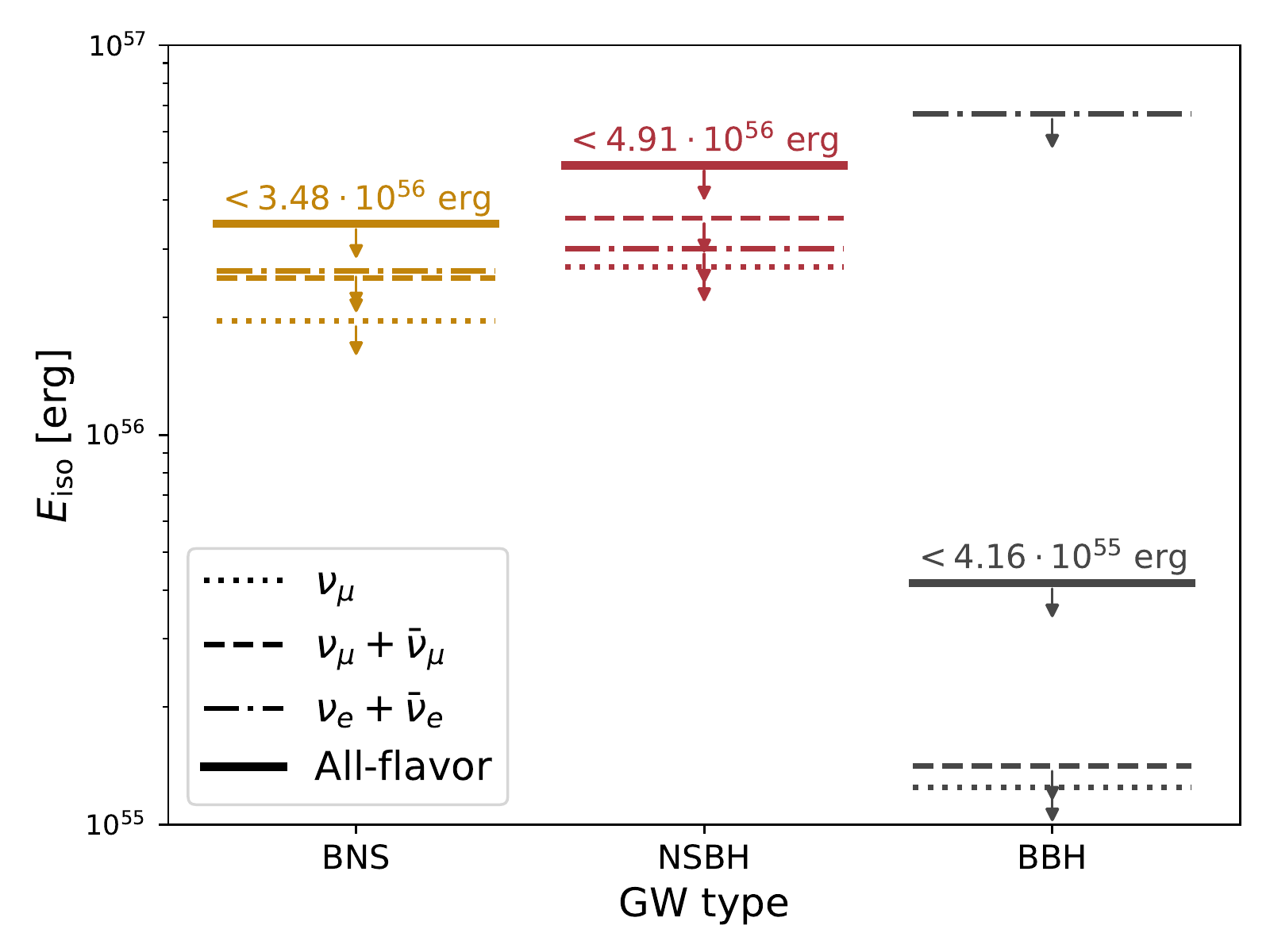}{0.48\textwidth}{(a) Limits on $E_{\rm iso}$ assuming same emission}
        \fig{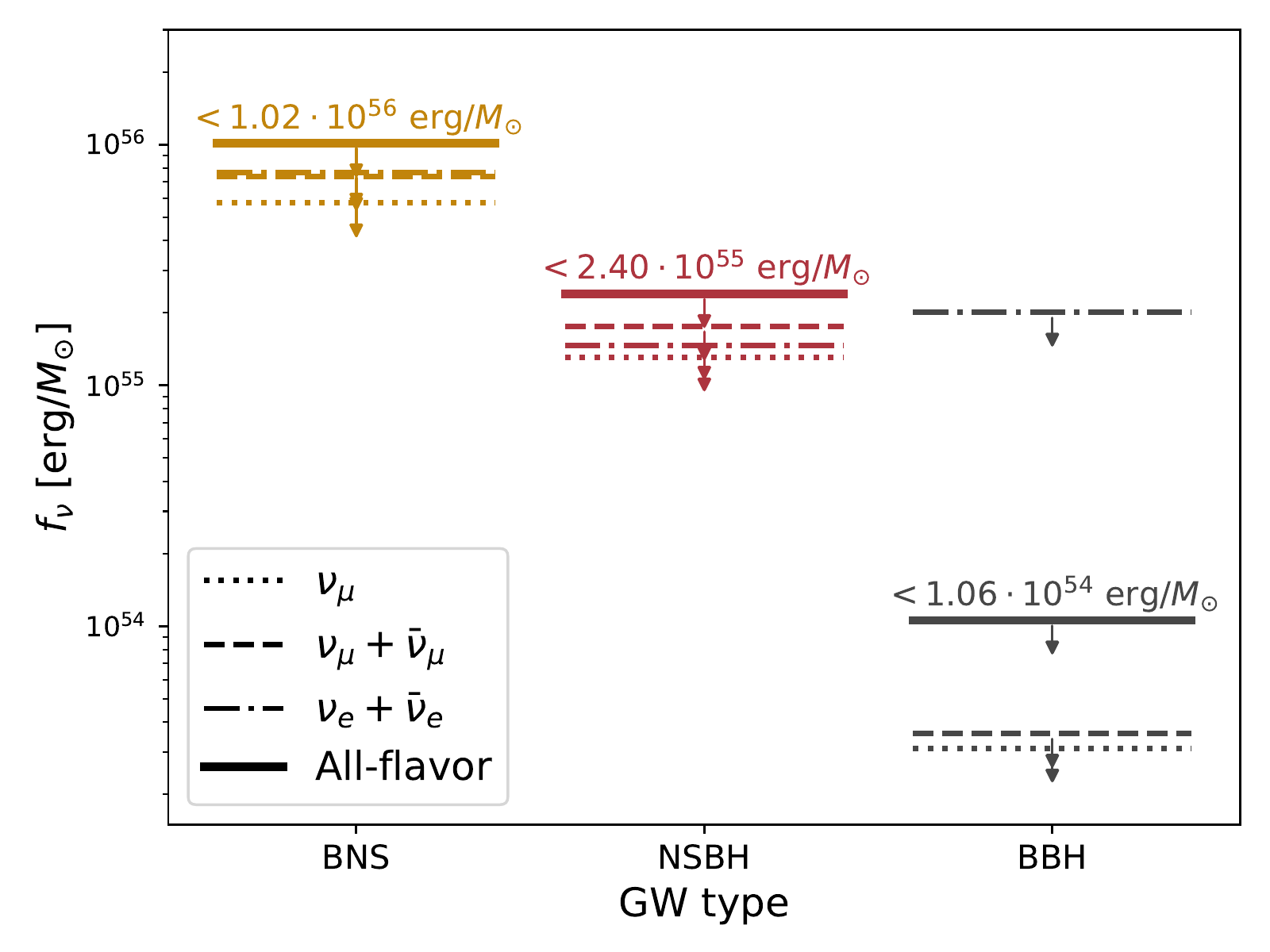}{0.48\textwidth}{(b) Limits on $f_\nu$ assuming scaling with total mass}
    }
    \caption{90\% C.L. upper limits on the isotropic energy emitted in neutrinos by combining GW triggers with the same nature, for $\nu_\mu$, $\nu_\mu+\bar\nu_\mu$, $\nu_e+\bar\nu_e$ and all-flavor emission (assuming equipartition). The left panel shows the results assuming that all selected sources are emitting the same $E_{\rm iso}$ while the right panel is assuming that neutrino emission is scaling with the total mass of the binary system.}
    \label{fig:Eiso_comb}
\end{figure}

\autoref{fig:Eiso_comb} presents the results for the three categories defined above: 1 BNS candidate\footnote{In this case, the result is the same as using directly GW190425 event, as it is the only identified BNS in O3a}, 2 NSBH (GW190426\_152155 and GW190814), 33 BBH (all other events in O3a). The all-flavor limit values are indicated on the figures, with the most constraining results obtained for the BBH population: $E_{\rm iso} < \SI{4.16e55}{\erg}$ assuming all objects have similar emission. This turns to $E_{\rm iso} < \SI{9.73e56}{\erg}$ for $E^{-3}$ spectrum.

Despite the objects being closer, the BNS and NSBH limits are worse than the ones for BBH because of the limited statistics for these two samples and of the fact that the three corresponding GW events have localization above the SK horizon.

\subsection{Low-energy neutrino emission}
\label{sec:popLE}

As for the flux limits, the low-energy case is much simpler. $E_{\rm iso}$ limits are directly obtained by scaling the flux limits using the source distance estimate. In case per-flavor limits are combined, the limit on the total energy emitted in all flavors, assuming equipartition, is, however, dominated by the $\bar\nu_e$ limit. To cover the distance uncertainty, the following likelihood was defined:
\begin{equation}
    \mathcal{L}(E_{\rm iso}; N_{\rm obs}, N_{\rm bkg}) = \displaystyle\int_0^{\infty} \dfrac{\left( N_{\rm bkg} + c^{\rm LE}(r) \times E_{\rm iso} \right)^{N_{\rm obs}}}{N_{\rm obs}!} e^{-\left( N_{\rm bkg} + c^{\rm LE}(r) \times E_{\rm iso} \right)} \times p_{\rm GW}(r) {\rm d}r,
\end{equation}
where $N_{\rm obs}$ and $N_{\rm bkg}$ are the observed and expected number of LE-$\nu$ events, $c^{\rm LE}(r)$ is the conversion factor from $E_{\rm iso}$ to number of signal events assuming Fermi-Dirac spectrum and source at distance $r$, $p_{\rm GW}(r)$ is the p.d.f. of distance estimation provided by LIGO-Virgo \citep{Singer:2016eax}. Detailed results for selected triggers are shown in \autoref{tab:flux}.

\section{Discussion and conclusions}
\label{sec:concl}

The results of the follow-up of LVC O3a gravitational waves with the SK detector have been presented. In the $\pm \SI{500}{\second}$ time windows centered on the triggers, no excess with respect to the background hypothesis was observed in any of the four considered samples (three for HE-$\nu$, one for LE-$\nu$). Upper limits on the incoming neutrino flux were computed for the different neutrino flavors. For HE-$\nu$, $E^{-2}$ spectrum was assumed, while for LE-$\nu$ limits, Fermi-Dirac emission with average energy of $\SI{20}{\mega\electronvolt}$ was considered. In both cases, detailed results are presented in the \autoref{tab:datarelease}. Assuming isotropic emissions and equipartition between the different flavors, upper limits on the total energy as neutrinos $E_{\rm iso}$ were derived, both individually for each trigger and by combining the different triggers of the same type, assuming the same emission or that the neutrino emission is scaling with the total mass of the binary system.

For low-energy neutrino emissions, the upper limits on the isotropic energy are not yet constraining enough to probe existing models such as \cite{Foucart:2015gaa} (predicted luminosity $L_{\rm iso}^{\rm model} \sim \SIrange{4}{7e53}{\erg\per\second}$), even though the exact shape of the neutrino spectrum (beyond the assumed simple Fermi-Dirac distribution with $\langle E_\nu \rangle = \SI{20}{\mega\electronvolt}$) may modify slightly the obtained upper limits.

For high-energy neutrino emissions, the obtained limits on $E_{\rm iso}$ assuming $E^{-2}$ spectrum are barely covering the nonphysical region where the total mass of the binary system is converted to neutrinos ($f_{\nu} \lesssim \SIrange[per-mode=symbol]{e54}{e56}{\erg\per\solarmass} \simeq 60-6000\%$), while the region currently probed by IceCube is $f_{\nu} \lesssim 1\%$)~\citep{Aartsen:2020mla}. However, this depends greatly on the assumed spectrum; if the latter happens to be different from the $E^{-2}$ standard scenario or features a cutoff, the limits would be changed as illustrated in \autoref{sec:popHE} for the $E^{-3}$ spectrum. A larger \si{\giga\electronvolt} component would favor detection and precise reconstruction of such neutrinos at SK as compared to larger neutrino telescopes like IceCube~\citep{Abbasi:2021kft}.

Even though the present paper has focused on the O3a catalog and the analysis was performed offline, the selections and techniques could also be used for real-time follow-up in the O4 observation period and beyond. With these constantly increasing statistics, it may finally be possible to probe the GW+$\nu$ source population and better understand the underlying mechanisms.

\begin{acknowledgments}
We gratefully acknowledge the cooperation of the Kamioka Mining and Smelting Company. The Super‐Kamiokande experiment has been built and operated from funding by the Japanese Ministry of Education, Culture, Sports, Science and Technology, the U.S. Department of Energy, and the U.S. National Science Foundation. Some of us have been supported by funds from the National Research Foundation of Korea NRF‐2009‐0083526 (KNRC) funded by the Ministry of Science, ICT, and Future Planning and the Ministry of Education (2018R1D1A3B07050696, 2018R1D1A1B07049158), the Japan Society for the Promotion of Science, the National Natural Science Foundation of China under Grants No.11620101004, the Spanish Ministry of Science, Universities and Innovation (grant PGC2018-099388-B-I00), the Natural Sciences and Engineering Research Council (NSERC) of Canada, the Scinet and Westgrid consortia of Compute Canada, the National Science Centre, Poland (2015/18/E/ST2/00758), the Science and Technology Facilities Council (STFC) and GridPPP, UK, the European Union's Horizon 2020 Research and Innovation Programme under the Marie Sklodowska-Curie grant agreement no.754496, H2020-MSCA-RISE-2018 JENNIFER2 grant agreement no.822070, and H2020-MSCA-RISE-2019 SK2HK grant agreement no. 872549.
\end{acknowledgments}

\appendix

\section{Additional material and data release}

This section details all the results not presented in the main text of the paper. The \autoref{tab:datarelease} contains the number of observed and expected events in the different samples and for each follow-up, as well as computed flux and $E_{\rm iso}$ limits.

Additionally, the data release~\cite{ZenodoLink} contains the effective areas that have actually been involved in the computation of flux upper limits, as presented in \autoref{sec:fluxHE:simple}. These can be used to derive again the upper limits with a specific source position or a different spectrum.

\begin{deluxetable*}{clll}
\tablecaption{Content of the detailed data release table.\label{tab:datarelease}}
\tabletypesize{\scriptsize}
\tablehead{
    \colhead{Col.\#} & \colhead{Label} & \colhead{Unit} & \colhead{Description}
}
\startdata
1 & \texttt{GW\_NAME} & - & Name of the GW trigger \\
2 & \texttt{GW\_UTC} & - & UTC time of the trigger \\
3 & \texttt{GW\_SKYAREA90} & deg$^2$ & Surface of the 90\% containment of GW localization \\
4 & \texttt{GW\_DISTANCE} & \si{\mega\parsec} & Mean estimate of the distance to GW source \\
5 & \texttt{SK\_LIVETIME} & \si{\second} & Live time of Super-Kamiokande over the selected \SI{1000}{\second} time window \\
6 & \texttt{SK\_FC\_OBSERVED} & - & Number of observed events in the HE-$\nu$/FC sample in the time window \\
7 & \texttt{SK\_FC\_EXPECTED} & - & Number of expected background events in the HE-$\nu$/FC sample in the time window \\
8 & \texttt{SK\_PC\_OBSERVED} & - & Same for HE-$\nu$/PC \\
9 & \texttt{SK\_PC\_EXPECTED} & - & Same for HE-$\nu$/PC \\
10 & \texttt{SK\_UPMU\_OBSERVED} & - & Same for HE-$\nu$/UPMU \\
11 & \texttt{SK\_UPMU\_EXPECTED} & - & Same for HE-$\nu$/UPMU \\
12 & \texttt{SK\_LOWE\_OBSERVED} & - & Same for LE-$\nu$ \\
13 & \texttt{SK\_LOWE\_EXPECTED} & - & Same for LE-$\nu$ \\
14 & \texttt{E2PHI90\_NUE\_FC} & \si{\giga\electronvolt\per\square\centi\meter} & 90\% U.L. on $E^2 \left. {\rm d}n/{\rm d}E \right\vert_{\nu_e}$ using the HE-$\nu$/FC sample, assuming $E^{-2}$ spectrum \\
15 & \texttt{E2PHI90\_NUEB\_FC} & \si{\giga\electronvolt\per\square\centi\meter} & Same for $E^2 \left. {\rm d}n/{\rm d}E \right\vert_{\bar\nu_e}$ \\
16 & \texttt{E2PHI90\_NUENUEB\_FC} & \si{\giga\electronvolt\per\square\centi\meter} & Same for $E^2 \left. {\rm d}n/{\rm d}E \right\vert_{\nu_e+\bar\nu_e}$ \\
17 & \texttt{E2PHI90\_NUMU\_FC} & \si{\giga\electronvolt\per\square\centi\meter} & Same for $E^2 \left. {\rm d}n/{\rm d}E \right\vert_{\nu_\mu}$ \\
18 & \texttt{E2PHI90\_NUMUB\_FC} & \si{\giga\electronvolt\per\square\centi\meter} & Same for $E^2 \left. {\rm d}n/{\rm d}E \right\vert_{\bar\nu_\mu}$ \\
19 & \texttt{E2PHI90\_NUMUNUMUB\_FC} & \si{\giga\electronvolt\per\square\centi\meter} & Same for $E^2 \left. {\rm d}n/{\rm d}E \right\vert_{\nu_\mu+\bar\nu_\mu}$ \\
20 & \texttt{E2PHI90\_NUE\_PC} & \si{\giga\electronvolt\per\square\centi\meter} & 90\% U.L. on $E^2 \left. {\rm d}n/{\rm d}E \right\vert_{\nu_e}$ using the HE-$\nu$/PC sample, assuming $E^{-2}$ spectrum \\
21 & \texttt{E2PHI90\_NUEB\_PC} & \si{\giga\electronvolt\per\square\centi\meter} & Same for $E^2 \left. {\rm d}n/{\rm d}E \right\vert_{\bar\nu_e}$ \\
22 & \texttt{E2PHI90\_NUENUEB\_PC} & \si{\giga\electronvolt\per\square\centi\meter} & Same for $E^2 \left. {\rm d}n/{\rm d}E \right\vert_{\nu_e+\bar\nu_e}$ \\
23 & \texttt{E2PHI90\_NUMU\_PC} & \si{\giga\electronvolt\per\square\centi\meter} & Same for $E^2 \left. {\rm d}n/{\rm d}E \right\vert_{\nu_\mu}$ \\
24 & \texttt{E2PHI90\_NUMUB\_PC} & \si{\giga\electronvolt\per\square\centi\meter} & Same for $E^2 \left. {\rm d}n/{\rm d}E \right\vert_{\bar\nu_\mu}$ \\
25 & \texttt{E2PHI90\_NUMUNUMUB\_PC} & \si{\giga\electronvolt\per\square\centi\meter} & Same for $E^2 \left. {\rm d}n/{\rm d}E \right\vert_{\nu_\mu+\bar\nu_\mu}$ \\
26 & \texttt{E2PHI90\_NUMU\_UPMU} & \si{\giga\electronvolt\per\square\centi\meter} & 90\% U.L. on $E^2 \left. {\rm d}n/{\rm d}E \right\vert_{\nu_\mu}$ using the HE-$\nu$/UPMU sample, assuming $E^{-2}$ spectrum \\
27 & \texttt{E2PHI90\_NUMUB\_UPMU} & \si{\giga\electronvolt\per\square\centi\meter} & Same for $E^2 \left. {\rm d}n/{\rm d}E \right\vert_{\bar\nu_\mu}$ \\
28 & \texttt{E2PHI90\_NUMUNUMUB\_UPMU} & \si{\giga\electronvolt\per\square\centi\meter} & Same for $E^2 \left. {\rm d}n/{\rm d}E \right\vert_{\nu_\mu+\bar\nu_\mu}$ \\
29 & \texttt{E2PHI90\_NUE\_COMBINED} & \si{\giga\electronvolt\per\square\centi\meter} & 90\% U.L. on $E^2 \left. {\rm d}n/{\rm d}E \right\vert_{\nu_e}$ using all HE-$\nu$ samples, assuming $E^{-2}$ spectrum \\
30 & \texttt{E2PHI90\_NUEB\_COMBINED} & \si{\giga\electronvolt\per\square\centi\meter} & Same for $E^2 \left. {\rm d}n/{\rm d}E \right\vert_{\bar\nu_e}$ \\
31 & \texttt{E2PHI90\_NUENUEB\_COMBINED} & \si{\giga\electronvolt\per\square\centi\meter} & Same for $E^2 \left. {\rm d}n/{\rm d}E \right\vert_{\nu_e+\bar\nu_e}$ \\
32 & \texttt{E2PHI90\_NUMU\_COMBINED} & \si{\giga\electronvolt\per\square\centi\meter} & Same for $E^2 \left. {\rm d}n/{\rm d}E \right\vert_{\nu_\mu}$ \\
33 & \texttt{E2PHI90\_NUMUB\_COMBINED} & \si{\giga\electronvolt\per\square\centi\meter} & Same for $E^2 \left. {\rm d}n/{\rm d}E \right\vert_{\bar\nu_\mu}$ \\
34 & \texttt{E2PHI90\_NUMUNUMUB\_COMBINED} & \si{\giga\electronvolt\per\square\centi\meter} & Same for $E^2 \left. {\rm d}n/{\rm d}E \right\vert_{\nu_\mu+\bar\nu_\mu}$ \\
35 & \texttt{EISO90\_NUE\_COMBINED} & \si{\erg} & 90\% U.L. on $\left.E_{\rm iso}\right\vert_{\nu_e}$ using all HE-$\nu$ samples, assuming $E^{-2}$ spectrum \\
36 & \texttt{EISO90\_NUEB\_COMBINED} & \si{\erg} & Same for $\left.E_{\rm iso}\right\vert_{\bar\nu_e}$ \\
37 & \texttt{EISO90\_NUENUEB\_COMBINED} & \si{\erg} & Same for $\left.E_{\rm iso}\right\vert_{\nu_e+\bar\nu_e}$ \\
38 & \texttt{EISO90\_NUMU\_COMBINED} & \si{\erg} & Same for $\left.E_{\rm iso}\right\vert_{\nu_\mu}$ \\
39 & \texttt{EISO90\_NUMUB\_COMBINED} & \si{\erg} & Same for $\left.E_{\rm iso}\right\vert_{\bar\nu_\mu}$ \\
40 & \texttt{EISO90\_NUMUNUMUB\_COMBINED} & \si{\erg} & Same for $\left.E_{\rm iso}\right\vert_{\nu_\mu+\bar\nu_\mu}$ \\
41 & \texttt{EISO90\_ALL\_COMBINED} & \si{\erg} & Same for all-flavors $E_{\rm iso}$ (assuming equipartition between flavors) \\
42 & \texttt{EISO90\_ALL\_COMBINED\_GAMMA3} & \si{\erg} & Same but assuming $E^{-3}$ spectrum \\
43 & \texttt{FLUENCE90\_LOWE\_NUE\_FERMIDIRAC} & \si{\per\square\centi\meter} & 90\% U.L. on $\Phi_{\nu_e}$ using LE-$\nu$ sample, assuming Fermi-Dirac spectrum \\
44 & \texttt{FLUENCE90\_LOWE\_NUEB\_FERMIDIRAC} & \si{\per\square\centi\meter} & Same for $\Phi_{\bar\nu_e}$ \\
45 & \texttt{FLUENCE90\_LOWE\_NUX\_FERMIDIRAC} & \si{\per\square\centi\meter} & Same for $\Phi_{\nu_\mu+\nu_\tau}$ \\
46 & \texttt{FLUENCE90\_LOWE\_NUXB\_FERMIDIRAC} & \si{\per\square\centi\meter} & Same for $\Phi_{\bar\nu_\mu+\bar\nu_\tau}$ \\
47 & \texttt{FLUENCE90\_LOWE\_NUE\_FLAT} & \si{\per\square\centi\meter} & 90\% U.L. on $\Phi_{\nu_e}$ using LE-$\nu$ sample, assuming flat spectrum \\
48 & \texttt{FLUENCE90\_LOWE\_NUEB\_FLAT} & \si{\per\square\centi\meter} & Same for $\Phi_{\bar\nu_e}$ \\
49 & \texttt{FLUENCE90\_LOWE\_NUX\_FLAT} & \si{\per\square\centi\meter} & Same for $\Phi_{\nu_\mu+\nu_\tau}$ \\
50 & \texttt{FLUENCE90\_LOWE\_NUXB\_FLAT} & \si{\per\square\centi\meter} & Same for $\Phi_{\bar\nu_\mu+\bar\nu_\tau}$ \\
51 & \texttt{EISO90\_LOWE\_NUE\_FERMIDIRAC} & \si{\erg} & 90\% U.L. on $\left.E_{\rm iso}\right\vert_{\nu_e}$ using LE-$\nu$ sample, assuming Fermi-Dirac spectrum \\
52 & \texttt{EISO90\_LOWE\_NUEB\_FERMIDIRAC} & \si{\erg} & Same for $\left.E_{\rm iso}\right\vert_{\bar\nu_e}$ \\
53 & \texttt{EISO90\_LOWE\_NUX\_FERMIDIRAC} & \si{\erg} & Same for $\left.E_{\rm iso}\right\vert_{\nu_\mu+\nu_\tau}$ \\
54 & \texttt{EISO90\_LOWE\_NUXB\_FERMIDIRAC} & \si{\erg} & Same for $\left.E_{\rm iso}\right\vert_{\bar\nu_\mu+\bar\nu_\tau}$ \\
55 & \texttt{EISO90\_LOWE\_ALL\_FERMIDIRAC} & \si{\erg} & Same for all-flavors $E_{\rm iso}$ (assuming equipartition between flavors) \\
\enddata
\tablecomments{Only the description of the table is shown here. The complete table is available as MRT format in the online journal and as CSV format in \cite{ZenodoLink}.}
\end{deluxetable*}

\bibliography{references}
\bibliographystyle{aasjournal}

\listofchanges

\end{document}